\documentclass[]{lnudissertation} % showframe
\usepackage[T1]{fontenc}
\usepackage[utf8]{inputenc}
\usepackage[swedish,english]{babel}

\usepackage[backend=biber,style=numeric,maxcitenames=1,maxbibnames=99]{biblatex} % style=alphabetic
\usepackage{url}
\usepackage{amsmath}
\usepackage{amssymb}
\usepackage{mathtools}
\usepackage{bm}
\usepackage{mathabx}
\usepackage{xfrac}
\usepackage{enumerate}
\usepackage{csquotes}
\usepackage{interval}
\usepackage{graphicx}
\usepackage{caption}
\usepackage{booktabs}
\usepackage{setspace}
\usepackage{siunitx}
\usepackage{enumitem}
\usepackage{multirow}
\usepackage{soul}
\usepackage{tikz}
\usepackage{lettrine}
\usepackage{epigraph}
\usepackage{verbatim} % for multi-line comments
\usepackage{newunicodechar}
\usepackage[hang,multiple]{footmisc}
\usepackage{tabularx}
\usepackage{minitoc}
\usepackage{pdfpages}
\usepackage[shortcuts]{extdash}
\usepackage{appendix}
\usepackage[framemethod=TikZ]{mdframed}
\usepackage{booktabs}

\usepackage[absolute,overlay]{textpos}
\usepackage{pbox}

\newcommand{\overlaypagenote}[1][This page will be replaced by the publisher (for the books in LUD Series)]{%
\begin{textblock*}{165mm}(0mm,230mm)%
\centering%
\small% 
\emph{#1}%
\end{textblock*}%
}

\newcommand{\ignore}[1]{}

\hypersetup{
	hidelinks, % Hide the boxes around hyperlinks
	pdfinfo={
		Title={Quantifying Process Quality: The Role of Effective Organizational Learning in Software Evolution},
		Author={Sebastian Hönel},
		Subject={PhD Dissertation},
		Keywords={Software Size; Software Metrics; Commit Classification; Maintenance Activities; Software Quality; Process Quality; Project Management; Organizational Learning; Machine Learning; Visualization; Optimization},
  }
}

\intervalconfig{
    soft open fences
}

\makeatletter
\def\url@acsstyle{%
  \def\UrlBreaks{\do\/}%
  \def\UrlSpecials{%
    \do\~{\penalty\UrlBreakPenalty\mathchar`~}%
    \do\.{\penalty\UrlBreakPenalty\mathchar`.}%
    \do\,{\penalty\UrlBreakPenalty\mathchar`,}%
    \do\-{\penalty\UrlBreakPenalty\mathchar`-}%
    \do\_{\penalty\UrlBreakPenalty\mathchar`_}%
    \do\?{\penalty\UrlBreakPenalty\mathchar`?}%
    \do\#{\penalty\UrlBreakPenalty\mathchar"23}%
    \do\%{\Url@percent}%
    \do\={\penalty\UrlBreakPenalty\mathchar`=\penalty\UrlBreakPenalty}%
    \do\&{\penalty\UrlBreakPenalty\mathchar`&\penalty\UrlBreakPenalty}%
    \do\ {\Url@space}\do\^^M{\Url@space}%
    \Url@force@Tilde}%
}
\makeatother

\usepackage{xurl}

\newcommand{\divergence}[2]{#1\,\|\,#2}
\newcommand{\abs}[1]{\left\lvert\,#1\,\right\rvert}

\DeclarePairedDelimiter\set\{\}
\newcommand{\tight}[1]{\,{#1}\,}
\newcommand{\utight}[1]{{#1}\,} % unary tight

\newcommand{\citepaper}[1]{Paper~\ref{app:publ-#1}}
\newcommand{\citepaperp}[1]{(\citepaper{#1})}
\newcommand{\citearticle}[1]{Article~\ref{app:publ-#1}}
\newcommand{\citearticlep}[1]{(\citearticle{#1})}

\DeclareUnicodeCharacter{0302}{aaaaaaaaaaaaaaaaaaaaaaaaaaaaaaaaaaaaaaaaaaaaaaa}

\begin{document}
%% Don't be surprised by *back*matter here, it is done to avoid page numbering, etc.
\backmatter

\begin{center}

%% Upper part of the page
\null\vspace{4.5cm}
\normalsize   Sebastian Hönel\\[1.5cm]
{ \Large \bfseries Quantifying Process Quality: The Role of Effective Organizational Learning in Software Evolution}%\\
%\large 
\\[2cm]

Doctoral Dissertation\\[0.2cm]
Computer Science / Software Technology\\[0.4cm]
2023
\vfill

%\draftstamp
%\draftstamp[PREPRINT]

%% Bottom of the page
%% The LNU logo image is copied from
%% https://medarbetare.lnu.se/medarbetare/stod-och-service/kommunikation-och-marknadsforing/designmanual/grundelement/logotyp/logotyp--symbol/
\includegraphics[scale=0.2]{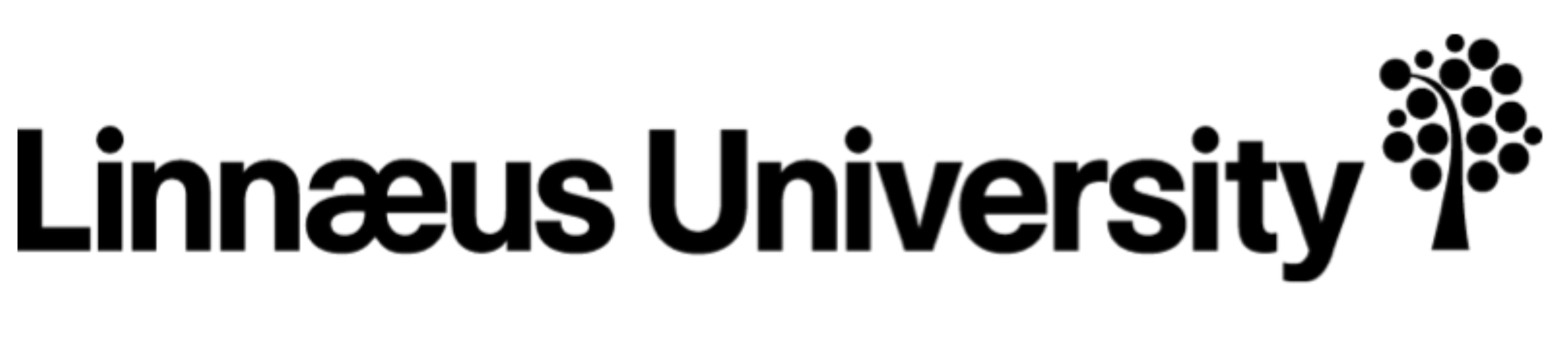}%

\end{center}

\overlaypagenote

\cleardoublepage

\null
\vfill
\noindent
\begin{minipage}{\textwidth}%
\raggedleft%
%\emph{Dedicated to\ldots}%
\emph{Dedicated to my father, who did not live to see me finish this.\\You are most heartfully missed.}
\end{minipage}%
\vfill

\cleardoublepage
\frontmatter
\begin{abstract}
%{\small
% Introduction, background
Real-world software applications must constantly evolve to remain relevant.
This evolution occurs when developing new applications or adapting existing ones to meet new requirements, make corrections, or incorporate future functionality.
Traditional methods of software quality control involve software quality models and continuous code inspection tools. These measures focus on directly assessing the quality of the software.
However, there is a strong correlation and causation between the quality of the development process and the resulting software product.
Therefore, improving the development process indirectly improves the software product, too.
To achieve this, effective learning from past processes is necessary, often embraced through post mortem organizational learning.
While qualitative evaluation of large artifacts is common, smaller quantitative changes captured by application lifecycle management are often overlooked.
In addition to software metrics, these smaller changes can reveal complex phenomena related to project culture and management.
Leveraging these changes can help detect and address such complex issues.

% Problem?
% Method, how did we try to solve the problem?
% Results/Analysis, what was the outcome?
% Discussion, what do the results mean?
Software evolution was previously measured by the size of changes, but the lack of consensus on a reliable and versatile quantification method prevents its use as a dependable metric.
Different size classifications fail to reliably describe the nature of evolution. While application lifecycle management data is rich, identifying which artifacts can model detrimental managerial practices remains uncertain.
Approaches such as simulation modeling, discrete events simulation, or Bayesian networks have only limited ability to exploit continuous-time process models of such phenomena.
Even worse, the accessibility and mechanistic insight into such gray- or black-box models are typically very low.
%Simulation modeling and Bayesian networks have limited ability to exploit continuous-time process models, with low accessibility and mechanistic insight.
To address these challenges, we suggest leveraging objectively captured digital artifacts from application lifecycle management, combined with qualitative analysis, for efficient organizational learning.
A new language-independent metric is proposed to robustly capture the size of changes, significantly improving the accuracy of change nature determination.
The classified changes are then used to explore, visualize, and suggest maintenance activities, enabling solid prediction of malpractice presence and -severity, even with limited data.
Finally, parts of the automatic quantitative analysis are made accessible, potentially replacing expert-based qualitative analysis in parts.
\bigskip
\vfill

\end{abstract}
\cleardoublepage

\begin{acknowledgments}
\null
A project like completing a Ph.D. cannot be accomplished without the support of others.
First and foremost I would like to express my sincere gratitude toward my supervisors Morgan, Welf, and Anna.
Only their unwavering support during the last years allowed me to explore the unknown and to become an autonomous researcher.
I would like to thank them for the countless meetings and immediate support whenever I needed it.

Getting children during this journey sure did not make things easier, although more pleasant.
This, however, is due to my life partner Frida, whose patience and structuredness allowed me to focus, even during times when time was scarce and pressure was high.
It would not have been possible without you, otherwise.
Thank you for taking care of me and our two wonderful children.
My family, even though they are far away, supported me in every way possible, too.
I am very grateful to my parents, Birgit and Norbert, and my sisters Christine and Tanja, and their families for frequently bridging the long-distance gap one way or another.

During the last few years, I did get to meet many very interesting people, many of who have become good friends.
I want to thank Petr and Premek for the intense, long, and good collaboration.
Maria and Rafael were early-on collaborators and I hope that there will be an opportunity to work together with them again.
I want to thank the Research School of Management and IT for accepting me as a member and allowing me to meet all the wonderful people there over the years.
I want to thank Mehdi and Farid, whom I share the office with for the good working climate, talks and laughs we continuously enjoy.
I would also like to acknowledge all the other Ph.D. students I met over the years, such as Angelos, Elizaveta, Kostiantyn, Manoranjan, Mohanraj, Nico, Olle, Zeynab, and many more.
There are many people working at the faculty that should be mentioned, too: Diana Unander, Diego Perez Palacin, Joachim Toft, Mathias Hedenborg, Tobias Andersson-Gidlund, and Romain Herault.
Thanks to Jonas Nordqvist for reading the first draft of my dissertation and giving me valuable feedback.

%% With the null, we avoid the space when starting a new paragraph, if necessary
%\null
\bigskip
\vfill
\noindent V\"axj\"o, Sweden\\
\noindent 2023\\
\bigskip

\end{acknowledgments}
\cleardoublepage

%% Start the ToC and add the corresponding lists
\renewcommand\contentsname{Table of Contents}
\setcounter{tocdepth}{1}

\phantomsection
\addcontentsline{toc}{chapter}{\contentsname{}}
%% To squeeze one more line of contents in the ToC page, if neccessary, modify the length as below:
%\addtolength{\cftaftertoctitleskip}{-11pt}
\tableofcontents
\cleardoublepage

%% With regard to figure and table sizes, the following could be used (figbox simply provides a gray frame):
%% wide figure: \figbox{\includegraphics[width=0.975\linewidth]{images/zzz.pdf}}
%% narrow figure: \figbox{\includegraphics[width=0.66\linewidth]{images/zzz.pdf}} (well, might be different than 0.66 depending on the context)
%% sideways figure: \begin{sidewaysfigure} ... \figbox{\includegraphics[width=0.985\linewidth]{images/zzz.pdf}} (make sure to check the margins with geometry package option 'showframe' !)
%% subfigures: \begin{subfigure}{\linewidth} ... \figbox{\includegraphics[width=0.975\linewidth]{images/zzz.pdf}} (make sure configure space between subfigures and captions properly)
%% tables: if required, \begin{table}[t] ... \begin{minipage}{0.975\textwidth} ... \begin{tabular}{lll}
%% Also, note that percent signs at the end of some commands/lines might be important to avoid extra space

\phantomsection
\addcontentsline{toc}{chapter}{\listfigurename{}}
\listoffigures
%\cleardoublepage
%
%\phantomsection
\addcontentsline{toc}{chapter}{\listtablename{}}
\listoftables
\cleardoublepage

%% The list of listings is not tested!!
%% Add the listings package in case it is needed, too
%\lstlistoflistings

% Add the list of publications
\newcommand{\printpublication}[1]{\AtNextCite{\defcounter{maxnames}{99}}\fullcite{#1}}
\newcommand{\myfullcite}[1]{\printpublication{#1}~\cite{#1}. {\footnotesize\textsc{In Appendix:}}~\ref{app:publ-#1}. Materials appear in Chapter}
\newcommand{\myfullcitedata}[1]{\printpublication{#1}~\cite{#1}. Materials used or appear in Chapter}
\newcommand{\myfullcitenoapp}[1]{\printpublication{#1}~\cite{#1}.}

\chapter*{List of Publications}
\addcontentsline{toc}{chapter}{List of Publications}
\markboth{List of Publications}{List of Publications}

\noindent\textbf{Canonical reference to this dissertation}
(the digital object identifier always points to the most recent master copy that receives corrections to errata found after print):

\begin{enumerate}[leftmargin=*]
    \item[]\printpublication{honel2023_phdthesis}~\cite{honel2023_phdthesis}.
\end{enumerate}

\bigskip
\bigskip
\noindent\textbf{This dissertation is based on the following refereed publications in chronological order} 
(I have contributed to all stages of work in the role of a lead author and have
implemented all tools):

\begin{enumerate}[series=publ,start=1,leftmargin=*]%[label=(\Alph*)]
    \item\myfullcite{honel2018changeset}~\ref{chap:obj-1}.
    \item\myfullcite{honel19importance}~\ref{chap:obj-1}.
    \item\myfullcite{honel2020using}~\ref{chap:obj-1}.
\end{enumerate}
\clearpage
\begin{enumerate}[resume*=publ]
    \item\myfullcite{picha_honel22pilot}~\ref{chap:obj-2}.
    \item\myfullcite{honel2022qrsmas}~\ref{chap:obj-3}.
    \item\myfullcite{honel2023hmm}~\ref{chap:obj-1}.
    \item\myfullcite{honel2023embedded}~\ref{chap:obj-2}.
    \item\myfullcite{honel2023masjoss}~\ref{chap:obj-3}.
\end{enumerate}

\newpage
\noindent\textbf{This dissertation is based on and produced the following replication packages, technical reports, posters, datasets, and software artifacts}
(I have contributed to these either exclusively or in the role of a lead author):

\begin{enumerate}[series=publ1,start=1,leftmargin=*]
    \item\myfullcitedata{honel2023dissrepl}s~\ref{chap:intro} through~\ref{chap:conclusions-future-work}.
    \item\myfullcitedata{honel2023fdtr}s~\ref{chap:obj-1},~\ref{chap:obj-2}, and~\ref{chap:obj-3}.
    \item\myfullcitedata{honel2018changeset_poster}~\ref{chap:obj-1}.
    \item\myfullcitedata{honel2019gitdens}~\ref{chap:obj-1}.
    \item\myfullcitedata{honel2019commits}~\ref{chap:obj-1}.
    \item\myfullcitedata{honel2023hmm_dataset}~\ref{chap:obj-1}.
    \item\myfullcitedata{honel2023fddataset}~\ref{chap:obj-2}.
\end{enumerate}
\clearpage
\begin{enumerate}[resume*=publ1]
    \item\myfullcitedata{honel2023fdgithub}~\ref{chap:obj-2}.
    \item\myfullcitedata{honel2023qcc}~\ref{chap:obj-3}.
    \item\myfullcitedata{honel2023masgithub}~\ref{chap:obj-3}.
\end{enumerate}

%\newpage
%\vfill
%\vspace{40pt}
\bigskip
\bigskip
\noindent\textbf{Further publications not related to this dissertation}
(I have contributed to all
or some stages of work in conceptualization, implementation, writing, or implementation of related tools):

\begin{enumerate}
    \item\myfullcitenoapp{ulan_honel2018qmio}
    \item\myfullcitenoapp{ulan_honel2018qmio_artifact}
    \item\myfullcitenoapp{honel2022pareto}
\end{enumerate}

\cleardoublepage

\cleardoublepage

%----------------------------------------------------------------------------------------------------
%---------------------HERE STARTS THE THESIS CONTENT-------------------------
%----------------------------------------------------------------------------------------------------
\mainmatter

%% Input chapters here
%\input{introduction}
\chapter{Introduction}\label{chap:intro}%
\vspace{5pt}
\chaptertoc

\begin{figure}[t!]
    \centering
    \includegraphics[width=.9\textwidth]{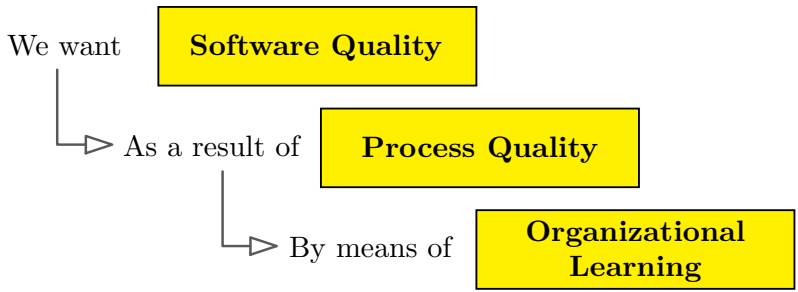}
    \caption[Illustration of the overall objective.]{{\small Improving software quality is the overall objective of this dissertation. The mean to that end is organizational learning from the outcomes of past projects' processes.}}
    \label{fig:overall-goal}
\end{figure}

\noindent
A common approach to measuring and improving software quality is the application of software quality models~\cite{Singh2013reviewqms}.
More recently, continuous code inspection tools are used as direct measures to advance software quality by means of identifying technical debt or predicting faults~\cite{alfayez2023sqtd,lomio2022sqfaults}.
Since software is the primary outcome of software processes~\cite{perkusich15procedure} and a strong causality between the quality of the process and the quality of the product exists~\cite{halvorsen2001}, the quality of the development process should be improved first and foremost.
There exist a multitude of facets that could be used to define process quality.
For example, some might consider time, effort/productivity, and defects (e.g.,~\cite{fenton1991metrics}), while others would emphasize stability and control (e.g.,~\cite{yeh1993procqual}).
Process quality may also be improved by, e.g., implementing process models or adopting standards.
Considering the development process as a single entity is perhaps too complex.
Therefore, it is suggested to decompose it into the elements and roles relevant for quality first~\cite{sorumgaard1995aspects}.
In this work, we focus on improving the quality of future processes primarily for the roles of developers and managers by learning from mistakes committed by them in the past.
This leverages the widely adopted practice of (iteratively) enhancing the process based on the results of post mortem (i.e., after project end) organizational learning (depicted in Figure~\ref{fig:overall-goal})~\cite{birk2002postmortem}.

In order to learn from a past project, its process needs to be quantifiable and captured.
During the complete lifecycle, a large number of changes are made to the software as it \emph{evolves}.
Software evolution happens when a software application is newly developed or adapted to new requirements, changed to accommodate new or changed functionality, or receives corrections for faults~\cite{bendifallah1987maintenance}.
The processes that guide software evolution are governed by technical artifacts that are required or produced during development.
Most of these artifacts can be summarized under \emph{application lifecycle management} data~\cite{chappell2008application}.
It contains, for example, technical documentation~\cite{souza2005documentation}, source code versioning systems, or bug trackers~\cite{antoniol2005vcs}.
The number of changes captured in the application lifecycle management data is usually considerable.
Observing numerous small changes should allow for capturing the process with the typical precision and detail required by statistical and machine learning.

It was previously recognized that it is of advantage to capture and comprehend the reasons for software change~\cite{mockus00identifying}.
Commonly, this knowledge is exploited for fault prediction and \mbox{-prevention}~\cite{graves2000predict}.
Beyond that, software evolution and especially its \emph{maintenance phases} lack a strong understanding, while simultaneously they consume a large proportion of the resources that are usually made available to a project~\cite{lientz1978maintenance}.
It is estimated that software maintenance even consumes up to half or more of the costs of the entire lifecycle of a software~\cite{boehm1976softweng}.
Some estimates even suggest that it consumes up to $100\times$ of the initial development costs~\cite{glinz2010}.
Therefore, it is worth understanding software maintenance and exploiting it for organizational learning by means of comprehensively evaluating past projects~\cite{dingsoyr2007postmortem}.

While the available application lifecycle management data is typically rich and high-dimensional, it is also often unstructured.
It is unclear how to automatically facilitate it for organizational learning, so engaging in it usually comes down to a manual, expert-based, qualitative analysis~\cite{picha2017towards}.
Expert-based assessment is subject to a variety of problems and factors affecting software-related estimates~\cite{Matsubara2021}.
What exacerbates this situation is that expert-based processes are often hard to repeat~\cite{macala96repeat}.
For these reasons, it is desirable to find a structured and automatic utilization of the available digital artifacts that can reduce the required manual effort for evaluating past projects.

We suggest a path towards automated and accessible organizational learning that is based on accurately capturing software size, evolution, and the nature of these changes (\textbf{O-I}), the identification and harnessing of apt application lifecycle management data for this task (\textbf{O-II}), and an enhancement of the explainability of decisions derived from quantitative data, making organizational learning more accessible to developers and managers (\textbf{O-III}).

\section{Objectives and Problems}\label{sec:objectives}
While post mortems represent a valid and proven measure of organizational learning~\cite{dingsoyr2007postmortem}, their conducting involves experts and the manual analysis of mostly unstructured process artifacts~\cite{muram2018,samalikova2014}.
This creates a variety of problems, such as discrepancies related to expert-judgment software estimates~\cite{Matsubara2021} or difficult-to-use process artifacts, which result in, e.g., hardly-repeatable evaluation processes~\cite{macala96repeat}.
The objectives of this dissertation, therefore, relate to the following overall problem:

\vspace{10pt}
\begin{mdframed}
    \vspace{7.5pt}
    How can we automatically facilitate, comprehend, and improve the accessibility of the available quantitative digital artifacts to improve process quality in order to reduce the required effort and dependency on experts for evaluating IT projects, such that they can be learned from later?
    \vspace{7.5pt}
\end{mdframed}
\vspace{10pt}

Organizational learning can be done on the basis of the results of a process quality evaluation.
As such, we require a device to reliably capture the process\,---\,the software evolution\,---\,in the first place.
While processes are often subdivided into larger cycles such as specification, implementation, or validation, we suggest capturing software evolution by larger quantities of small and tiny changes that are observable in the application lifecycle management data instead.
%such as (the size of) commits made to a repository or tickets opened in an issue-tracking system.
The overall problem was deconstructed into \textbf{three consecutive} objectives.
Each objective is accompanied by concrete problems that needed to be solved in order to achieve the objective.

The first objective, \textbf{O-I}, was identified after recognizing an existing research gap that no reliable method (that is, cross-language and cross-application) for quantifying software size, exists (e.g.,~\cite{herraiz06comparison}).
The size of software is among the most frequently used predictors for cost, effort, and productivity~\cite{chen2005costmodeling,kitchenham2004softwareprod}.
The lack of a consensus on what constitutes an adequate implementation for software sizes may even prohibit its usage for the aforementioned scenarios.
Therefore, a robust and versatile definition for software size was required.
While software size is an important predictor for effort, we suggest it can also be used to classify changes, that is, inferring certain types of activities that are carried out during the development process.
This kind of classification was attempted previously using more complex (or expensive-to-compute) features, such as leveraging semantic changes or inspection of the abstract syntax tree~\cite{jacksonL94semdiff,fluri07changedistill}.
Having a reliable classifier for changes (the \emph{evolution} of software) allows capturing otherwise latent properties of the process, such as software aging~\cite{fluri06classifying} or even behavioral patterns~\cite{fluri08discovering,levin16devprofile}.
We suggest leveraging software change for aspects of software process quality.

\vspace{5pt}
\begin{mdframed}[nobreak]
    \vspace{5pt}
    \begin{enumerate}[noitemsep,leftmargin=2em]
        \item[\textbf{O\=/I}:] Establish an efficient and effective size-based change detection that can be used for capturing the integral aspects and nature of software evolutionary processes.\begin{enumerate}[noitemsep, leftmargin=1.25em]
            \item[\textbf{P\=/I:}] Find a robust, versatile, and effective metric for measuring software size.
            \item[\textbf{P\=/II:}] Obtain a reliable and accurate change classifier that generalizes beyond single projects and/or languages.
        \end{enumerate}
    \end{enumerate}
    \vspace{2.5pt}
\end{mdframed}
\vspace{5pt}

The second objective, \textbf{O-II}, relies on a robust change classifier.
For this objective, we aim at exploiting application lifecycle management data for detecting problems in the software process.
Application lifecycle management data provides a rich set of digital artifacts, some of which are structured data~\cite{chappell2008application}.
For example, it contains the source code repositories and commits thereof, which are small changesets that have been previously used to predict the nature of the change~\cite{levin17boosting}.
Another example is data from issue-tracking systems (e.g., bug trackers or project management software).
We suggest that such data can be leveraged to model \emph{activities} that describe the software process well, similar to the so-called disciplines in the Unified Process~\cite{Scott2001uniproc}.
We hypothesize that some of these activities, while derived from data governing the development process, also reflect actions carried out by the management, which can be exploited to detect a class of problems called project management \emph{anti-patterns}~\cite{picha2019apdetect,brada2019catalogue}.
While the detection of such problems has been attempted previously, those attempts were limited in nature leaving open a research gap for more holistic approaches.

\vspace{5pt}
\begin{mdframed}[nobreak]
    \vspace{5pt}
    \begin{enumerate}[noitemsep,leftmargin=2.5em]
        \item[\textbf{O\=/II}:] Enable the harnessing of application lifecycle management data for aspects of software process quality and organizational learning thereof.\begin{enumerate}[noitemsep, leftmargin=1.25em]
            \item[\textbf{P\=/III:}] Identify whether, which, and how application lifecycle management data can be leveraged for process modeling.
            \item[\textbf{P\=/IV}:] Obtain a stable predictive model for assessing anti-pattern presence and -severity.
            \item[\textbf{P\=/V}:] Find a replicable methodological framework that enables automatic organizational learning from a few past software\\projects.
        \end{enumerate}
    \end{enumerate}
    \vspace{2.5pt}
\end{mdframed}
\vspace{5pt}

The third and last objective, \textbf{O-III}, is another direct continuation of our efforts related to the previous objective.
At this point, we will have addressed problems three through five.
However, the last objective is concerned with making organizational learning more accessible by moving away from qualitative analysis towards automatic, quantitative analysis, as the former often requires experts who are often unavailable, expensive, or introduce their own subjective bias~\cite{Matsubara2021,muram2018}.
We suggest achieving the last objective by enhancing the explainability of quantitative results through the usage of \emph{scores}, a mathematical construct that allows arbitrary numeric quantities to be compared, comprehended, and associated with some notion of quality in a straightforward way.
The conjunction of scores with variable importance~\cite{zhu2015varimp} may simplify the understanding of certain facets of problems related to software processes.
With a new approach called \emph{Automatic Calibration}, we suggest a largely unsupervised technique to go towards ``white-box''-models, which\,---\,compared to ``black-box''-models\,---\,provide traceability and transparency of their inner components and mechanics~\cite{Rai2019xai}.

\vspace{5pt}
\begin{mdframed}[nobreak]
    \vspace{5pt}
    \begin{enumerate}[noitemsep,leftmargin=3em]
        \item[\textbf{O\=/III}:] Enhance the explainability of quantitative data and reduce the amount of required qualitative analysis.\begin{enumerate}[noitemsep, leftmargin=1.25em]
            \item[\textbf{P\=/VI:}] Find a technique for raw quantities that allows them to be understood, compared, and associated with quality.
            \item[\textbf{P\=/VII:}] Suggest a solution that can provide details about phenomenon severity and serve as a data-based drop-in replacement for qualitative evaluation.
        \end{enumerate}
    \end{enumerate}
    \vspace{2.5pt}
\end{mdframed}

\section{Contributions}
The works presented in this dissertation claim the following contributions.
Included publications are referred to by (conference) \emph{Paper} and (journal) \emph{Article} and are to be found in dedicated appendices.

\begin{enumerate}[series=contrib,start=1,leftmargin=*,label=(\roman*)]
    \item A well-defined and versatile new software metric, the source code density, is suggested. The metric is a computationally cheap and language-independent (i.e., applicable to a variety of programming languages) quantifier for software size~\cite{honel2018changeset_poster}.
    \item A promising approach to assessing developer efficacy that is based on source code density and time spent as a notion of effort~\citepaperp{honel2018changeset}.
    \item A reliable cross-project classifier for the nature of changesets that largely derives its predictions from source code density (\citepaper{honel19importance} and \citearticle{honel2020using}). An additional considerable improvement to this classifier is suggested based on relations and sojourn-times between changesets~\citepaperp{honel2023hmm}.
    \item Software, datasets, and experimental setups related to source code density that can be used to extract the new metric, conduct analyses, and replicate our work~\cite{honel2019gitdens,honel2019commits,honel2023hmm_dataset}.
    \item A pilot case study that examines which and how application lifecycle management data can be leveraged for the data- and rule-based detection of so-called anti-patterns~\citepaperp{picha_honel22pilot}.
    \item A full-scale embedded and longitudinal case study of $n\tight{=}15$ software projects that are affected by a Fire Drill to some degree. The case study suggests a general framework for replicating it in similar contexts~\citearticlep{honel2023embedded}.
    \item Software, datasets, and experimental setups related to the detection and severity assessment of managerial anti-patterns. These can be used to replicate our results~\cite{honel2023fddataset,honel2023fdgithub}.
    \item A generalized and universal approach of transforming arbitrary numeric quantities into \emph{scores}, a mathematical construct for equating otherwise incomparable quantities, such as software metrics~\citepaperp{honel2022qrsmas}.
\end{enumerate}
\clearpage
\begin{enumerate}[resume*=contrib]
    \item A utilization of this approach in the form of a visual analytics application. The interactive tool is implemented as a web application, supports generic datasets, and facilitates a wide range of statistical tests and the fitting of $\utight{\approx}120$ parametric probability distributions. It automatically generates summary statistics and scientific reports thereof~\citearticlep{honel2023masjoss}.
\end{enumerate}

\section{Thesis Outline}
The remainder of this dissertation is structured as follows.
Each of the Chapters~\ref{chap:obj-1},~\ref{chap:obj-2}, and~\ref{chap:obj-3} is dedicated to a single objective, its related problems, and suggested solutions.
The organization of these chapters follows the same structure.
First, the contributions are rehashed, before the objective and related problems are introduced.
Then, some background information is outlined.
Lastly, the publications that provide a solution to the presented problems are summarized in greater detail.
%
%in which the objective and problems are introduced first, background information is given, the list of related publications is shown, and the suggested solutions
In Chapter~\ref{chap:obj-1}, an efficient and effective size-based change detection utility is established.
The work presented in Chapter~\ref{chap:obj-2} exploits this mechanism for phenomena that continuously unfold over time for organizational learning.
Chapter~\ref{chap:obj-3} then is concerned with increasing the accessibility of organizational learning that is mainly based on quantitative data.
Chapter~\ref{chap:conclusions-future-work} summarizes and concludes this dissertation, giving closing remarks as well as prospective directions for future work.
Lastly, the publications included in this dissertation are to be found in the Appendix.

\cleardoublepage
\chapter{%
%Efficient and Effective
Size-Based Change Detection During Software Evolution}\label{chap:obj-1}%
\vspace{5pt}
\chaptertoc
\chaptermark{Size-Based Change Detection}
%
%\vspace{20pt}
%\clearpage
%\section{Overview and Objective}
%
\noindent
The contributions of this chapter are mainly based on the four refereed publications \citepaper{honel2018changeset}, \citepaper{honel19importance}, \citearticle{honel2020using}, and \citepaper{honel2023hmm}, summarized in Sections~\ref{ssec:summary-honel2018changeset}, \ref{ssec:summary-honel19importance}, \ref{ssec:summary-honel2020using}, and~\ref{ssec:summary-honel2023hmm}, respectively.
In \textbf{\citepaper{honel2018changeset}}, approximately $80,000$ commits from $1,650$ open source projects are analyzed to determine source code density.
The study reveals insights into commit behavior but cannot accurately predict effort.
Clone detection and string similarities are deemed unnecessary, and findings on software evolution apply beyond open source projects.
\textbf{\citepaper{honel19importance}} is our first study to focus on commit classification.
Its objectives are to analyze commit sizes, replicate state-of-the-art models, test source code density, and challenge existing findings.
The results revealed that commit size impacts commit nature, and net notions affect file and line counts.
The state-of-the-art model achieved approximately $76$\% accuracy, which we successfully replicate.
The source code density and size-based features prove to be viable replacements for semantic code changes without compromising performance.
We reliably boost classification accuracy to $80$\%, peaking at approximately $89$\%.
\textbf{\citearticle{honel2020using}} extends the previous study on commit classification, proposing that information from preceding commits can enhance classification performance.
By predicting the nature of the youngest commit based on its predecessors, the study considers maintenance activities and their probabilities, creating a dataset with consecutively labeled commits.
The results show that using all available features and considering up to three previous commits significantly improves accuracy, achieving over $93$\% accuracy, together with almost perfect Kappa values.
\textbf{\citepaper{honel2023hmm}} explores classifying commits into maintenance activities using various models, including hidden Markov models, dependent mixture models, and joint conditional density models.
The research aims to determine whether stateful models outperform stateless models and assess the performance of different models.
The results show that stateless joint conditional density models outperform other classifiers.
The study concludes that relational properties and sojourn times between commits are valuable for prediction and emphasizes the need for a larger training set and validation for a practical solution.

\section{Objective and Problems}
%
%\vspace{10pt}
%\noindent
This chapter is dedicated to the first objective (see Section~\ref{sec:objectives}).
It summarizes the results of the four refereed publications that we previously recapped.
This chapter's objective and its related problems are defined as follows.

\vspace{10pt}
\begin{mdframed}
    \textbf{O-I:}~\textit{Establish an efficient and effective size-based change detection that can be used for capturing the integral aspects and nature of software evolutionary processes.}
\end{mdframed}
\vspace{5pt}

\noindent In order to achieve the objective, the following problems need to be solved:

\begin{enumerate}[itemsep=.5ex, leftmargin=3.5em]
    \item[\textbf{P\=/I:}] Find a robust, versatile, and effective metric for measuring software size.
    \item[\textbf{P\=/II:}] Obtain a reliable and accurate change classifier that generalizes beyond single projects and/or languages.
\end{enumerate}

\section{Software Size}
%
%\lettrine[findent=2pt]{\fbox{\textbf{S}}}{ }
Software size has a history of being utilized for, e.g., cost modeling~\cite{chen2005costmodeling} or productivity- and effort estimation~\cite{kitchenham2004softwareprod}.
This was likely inspired by other manufacturing industries, in which the size of a product positively correlates with the time it took to fabricate it and the resources required for doing so.
However, unlike other industries, software size is not necessarily a straightforward measure for the cost of time and resources and might even be misleading if used inappropriately~\cite{kitchenham2007misleading}.
There are attempts to replace plain size metrics with more accurate reflections of the effort that went into implementing a certain functionality, such as function points, differencing algorithms, or the quantification of semantic changes~\cite{behrens83funcpoints,jacksonL94semdiff}.
These alternative attempts to software size, while working, have various drawbacks.
For example, function points have a strong positive correlation with lines of code, but are more difficult to compute~\cite{albrecht83funcpoints}.
Some differencing algorithms use the abstract syntax tree, which requires a program that can be compiled and is potentially computationally expensive as a static code analysis~\cite{fluri07changedistill}.

Many software metrics are counts, ratios, or aggregations thereof~\cite{carleton1999measure}.
Some of these metrics, for example, count the number of statements or occurrences of events or aggregate other metrics into higher-level quality goals~\cite{fitzpatrick00abc}.
This is similarly the case for software size when it is implemented as a variant of a count-based software metric (e.g., \emph{Lines Of Code}, \emph{Number Of Files}, etc.).
Since it can often be calculated with minimal computational effort, it is a desirable software metric.
However, it does not appear that there is a clear consensus on how to count the lines of code and derive a corresponding metric (e.g.,~\cite{herraiz06comparison}).
More attempts were made to further exploit software size as a predictor for various quantities, such as fault frequencies.
Most frequently, some form of lines of code is used as a predictor for the risk associated with changing software~\cite{mockus00risk}, the prediction and localization of faults~\cite{leszak02classification,purushothaman05small}, or for \emph{classifying changes}~\cite{hindle08whatdo}.

\section{Classifying Changes}
%
%\lettrine[findent=2pt]{\fbox{\textbf{M}}}{ }
More narrow applications of software size, such as fault prediction or localization, exist.
However, being able to classify the changes that were made to a software allows us to capture integral aspects of software evolution, such as the kind of work that was done or even behavioral patterns~\cite{fluri08discovering} (e.g., developer maintenance profiles~\cite{levin16devprofile}).
Understanding the nature or even the reason for change during software evolution can be leveraged as a measure to alleviate or mitigate \emph{software aging}~\cite{fluri06classifying}.
Changes indicate that the process alternates between maintenance phases.
Different maintenance phases commonly overlap in modern projects, as various activities are carried out simultaneously, often by multiple developers.
Being able to gain insight into these phases is valuable since an estimated $60$\% of maintenance work is spent on comprehension~\cite{kuhn07comprehension,abran05swebok}.
We have previously suggested that the ability to understand change may also be exploited for software quality monitoring or \emph{process pattern detection}~\citepaperp{honel19importance}.
%~\cite{honel19importance}.
Therefore, it is all the more desirable to classify changes using computationally cheap metrics based on software size.

% What is a change?
Modern software development facilitates version control systems to manage the source code of applications in development.
Historically relevant systems that track changes are, for example, Apache Subversion, Mercurial, or Git.
These systems exist to allow multiple developers to make changes in a concurrent and independent fashion and to track multiple revisions of an application simultaneously.
In the most popular version control system, Git, the source code is held in what is called a \emph{repository}, and changes are recorded as what are called \emph{commits}.
A repository is an archive representing the full history of an application, as it logs all \emph{incremental} changes that were made since the beginning.
Commits are often also called \emph{changesets}, because a single commit can comprise any number and type of changes, such as modified, added, or deleted files.
For a repository $\mathbb{R}_T$ of $t\in1\ldots n$ discrete and incremental changes, a commit $C_t$ represents the \emph{patch} between two consecutive states $R_{t-1},R_t$ of the same repository, such that $R_t=C_t\left(R_{t-1}\right)$.
The first commit is often called the \emph{initial} commit, since no previous changes existed prior to it, i.e., $R_{t=1}=C_{t=1}\left(\varnothing\right)$.
The initial commit, therefore, cannot contain modified, renamed, or deleted files, but only additions.

% What features a commonly used for classification?
Because of their nature, commits are the most commonly used unit for encapsulating and exploring changes (e.g.,~\cite{mockus00identifying,hindle09automatic}).
In addition to the set of changes they represent, commits are associated with a rich set of metadata, such as messages, keywords, names and email addresses of authors and committers, tags, timestamps, branches, or parent- and child-commits (e.g.,~\citepaper{honel2023hmm}).
Therefore, commits can reflect \emph{software evolution} from a source code perspective.
Metadata, especially commit messages and keywords, are frequently used for predicting the nature of a commit, as these may give a direct indication as to the purpose of the change (e.g.,~\cite{levin17boosting,mockus00identifying}).
However, these metadata are subject to human error; inaccurate, incomplete, and sometimes completely missing.
For that reason, others have attempted to derive a more robust classification that is based on the (semantic) changes a commit comprises~\cite{jacksonL94semdiff,fluri06classifying,fluri07changedistill}.
Finally, a number of studies focus primarily or exclusively on features derived from some notion(s) of size, such as the number of files or lines of code (e.g.,~\cite{hindle08whatdo,purushothaman05small,herraiz06comparison,hattori08nature}).

% Maintenance activities?
The most widely accepted categorization of evolutionary changes, \textbf{maintenance activities}, are \emph{adaptive}, \emph{corrective}, and \emph{perfective}~\cite{swanson76acp}.
Some studies, however, introduce additional categories or subdivide existing ones \cite{lin88class,hindle09automatic}.
Adaptive changes are forward engineering, that is, adding new features to an application.
Corrective changes are related to fixing bugs and correcting faulty or unintended behavior (both functional and non-functional).
Perfective changes, refactorings, and restructurings are made for maintenance reasons, such as accommodating future features, improving performance, or redesigning the system.
They do not alter the application's \emph{external} behavior.
Perfective changes are also the de facto way to reduce technical debt~\cite{avgeriou16techdebt}.

Each of the following sections addresses one of the problems and then summarizes the publications related to it that are included as part of this dissertation.

\section{Source Code Density}
The first problem (\textbf{P\=/I}) that needed to be addressed was to design a robust, versatile, and size-based metric because there was no consensus on how to measure software sizes definitively (e.g.,~\cite{herraiz06comparison}).
As we desire a computationally cheap software metric based on size, we suggest \emph{Source Code Density} as a more accurate reflection of software size (\citepaper{honel2018changeset} and~\cite{honel2018changeset_poster}).
%~\cite{honel2018changeset,honel2018changeset_poster}.
It is a metric that expresses the ratio between \emph{gross} and \emph{net} size which can be applied to an entire application or even single commits.
The gross size is the raw count of lines of code that includes duplicated (type-IV semantic clones) and dead code (unreachable or never called), comments, whitespace, and other (string-)similarities.
The net size expressly excludes all of these.
Hence the resulting ratio has a range of $\interval{0}{1}$, where a value of $0$ indicates that none of the code has a functional meaning (or that the gross size was zero), and a value of $1$ indicates that all the code contributes to the functionality.
Type-IV clones detect two or more code fragments that perform the same computation but are implemented by different syntactic variants.
Therefore, type-IV clones are resilient to, e.g., differing variable names, white space, etc.~\cite{roy09clones}.
Type-IV clone detection is the most sophisticated and exceeds the capabilities of types I-III, which are limited to exact copies (type-I), syntactically identical copies (type-II), and limited modifications, such as changed statements (type-III)~\cite{koschke06clones}.
The elements that go into the source code density are rather language-independent and cheap to compute (except for the clone detection and string similarities).
However, different programming languages will have different typical densities.
Moreover, two semantically equal implementations of the same functionality can often be implemented in a great number of ways.
For example, a multiple-line \textsf{for}-loop could be unrolled or expressed in perhaps just a single line using functional programming.

\begin{figure}[t!]
    \centering
    \includegraphics[width=\textwidth]{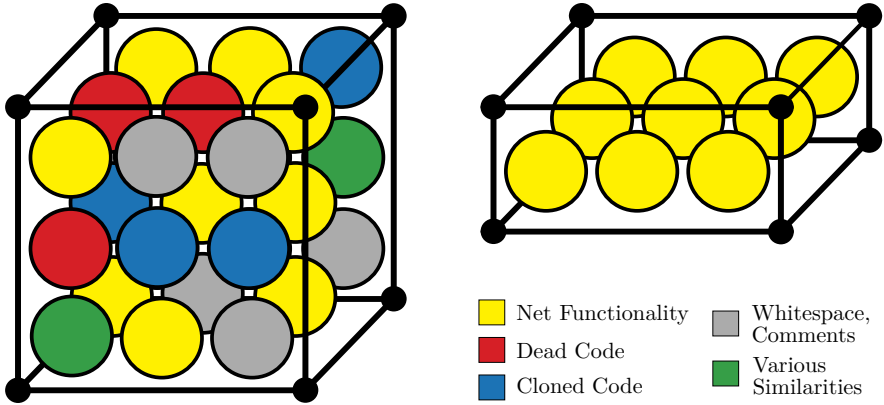}
    \caption[Gross size versus net functionality of a commit.]{{\small Illustration of a low-density ($\utight{\approx}\sfrac{1}{3}$) commit and its gross size (left), with various irrelevant elements. On the right side, the commit is reduced to its net functionality.}}
    \label{fig:low-high-dens}
\end{figure}

The source code density as a new software metric can be extracted using \emph{Git Density}~\cite{honel2019gitdens}, a tool created for this purpose.
It also extracts commit metadata, as well as approximately more than $20$ size-based other features, such as the number of modified files or the number of lines added by renamed files (net/gross).
In addition to that, it counts $20$ specific keywords, such as ``fix'', ``implement'', etc., as suggested by~\citeauthor{levin17boosting}~\cite{levin17boosting}.
Git Density was used to produce a dataset of more than $359,000$ commits, all of which have extensive size attributes attached~\cite{honel2019commits}.
Table~\ref{tab:dens-percentiles} shows the percentiles for the source code density of the commits in this dataset.
The average density is $\utight{\approx}0.81$ (see also Figure~\ref{fig:low-high-dens}).

%\subsection*{Study:~\citefield{honel2018changeset}{title}}
\subsection{Effort Estimation Using Changesets}\label{ssec:summary-honel2018changeset}
\paragraph{Summary of~\citepaper{honel2018changeset}.}
Here, we compute the source code density of $\utight{\approx}80,000$ commits for $1,650$ open source projects, about $800$/$400$/$400$ of which are primarily written in Java, C\#, and PHP.
%We attempt to predict a notion of effort by putting source code density into relation to the time spent, on a per-commit basis.
More accurate notions of the size of a commit, such as its source code density, can be facilitated for effort- and productivity estimations.
The International Software Benchmarking Standards Group (ISBSG)\footnote{ISBSG: ``Software size as the main input parameter to cost estimation models.'' 2018.~\href{https://web.archive.org/web/20180212115016/https:/www.isbsg.org/software-size/}{http://isbsg.org/software-size/}.} identifies software size as the main input to cost estimation models.
Size, however, is not a well-defined standard.
The notion of effort that we seek to predict is the size over time, and we refine it to (the amount of) \emph{net functionality} over time.
This study, therefore, examines the relationship between the density of a commit and how much time was spent on creating it.
The amount of time spent on a commit was obtained using ``Git hours''\footnote{Git hours: ``Estimate time spent on a git repository.'' 2018.~\href{https://web.archive.org/web/20180611015726/https://github.com/kimmobrunfeldt/git-hours}{https://github.com/kimmobrunfeldt/git-hours}.}.

\begin{table}[t!]
    \centering
    \resizebox{\textwidth}{!}{%
    \begin{tabular}{l|r|r|r|r|r|r|r|r|r|r}
        Percentile  & $0$\%   & $6.5$\%   & $10$\%    & $20$\%    & $30$\%    & $40$\%    & $50$\%    & $60$\%    & $70$\%    & $72.5$\% \\ \hline
        Density     & $0.0$   & $0.5$     & $0.57$    & $0.68$    & $0.75$    & $0.81$    & $0.86$    & $0.91$    & $0.97$    & $1.0$
    \end{tabular}
    }
    \caption[Typical percentiles for the source code density of more than $359,000$ commits.]{{\small Some percentiles for the source code density in a dataset of more than $359,000$ commits~\cite{honel2023dissrepl}. Approx. $27.5$\% of the commits have a density of $1.0$ and less than $6.5$\% of all commits have a density of $0.5$ or less, suggesting that the majority of commits have a high density. The expected source code density in this dataset is $\utight{\approx}0.81$~\cite{honel2019commits}.}}
    \label{tab:dens-percentiles}
\end{table}

While the results do show statistically significant correlations between these two, we cannot conclude that predicting the effort would work well, as the notion of time spent was automatically derived and is too coarse to yield any meaningful results.
However, we gain a few other valuable insights from this study.
Most importantly, clone detection is not necessary for obtaining precise results, that is, it does not have a significant impact on correlations with the time spent.
Moreover, computing string similarities between code fragments do not significantly improve the precision of the notion of source code density and can, therefore, be dispensed with as well.
We also got a first glimpse at detecting behavioral patterns using changes:
We were able to unveil a typical and reoccurring behavior of developers, in which they would make commits typically within $30$ minutes after they started their session, more commonly even after only a few minutes, and sometimes within less than a minute.
This is a phenomenon known as \emph{bulk-committing}, as developers tend to accumulate work only to then perform multiple consecutive commits within a relatively short period of time.
These ``micro''-patterns are commonly found.
Another example is that of impure commits, in which a commit \emph{tangles} unrelated changes.
This then leads to commits that are not purely assignable to any maintenance activity~\cite{kirinuki14hey}.

In the studies related to software size, we exclusively make use of \emph{libre} software, that is, open source software that is ubiquitously available and freely accessible on platforms such as GitHub\footnote{The GitHub Open Source collaborative platform:~\url{https://github.com/}.}.
While~\citeauthor{lehman80laws}'s laws~\cite{lehman80laws} suggested early on that libre and closed/proprietary software evolve differently, more recent studies find evidence that is counter-indicative of this fact.
For example,~\citeauthor{herraiz06comparison}~\cite{herraiz06comparison} have examined common evolutionary patterns across a large number of projects using size-based metrics.
They report that there are no significant differences in how these patterns manifest.
This suggests that our findings have some validity beyond libre software as well.

\section{Commit Classification}
An integral aspect of software evolution is being able to reason about, e.g., the quality of the process or product.
If such properties are derived from commits that were previously classified into maintenance activities~\cite{swanson76acp}, then the accuracy of this classification directly impacts the quality and validity of the conclusions.
A wide selection of repository artifacts, such as bug reports, issues, or code changes, have been used for the detection of refactoring types.
However, commits are the most widespread artifact used as a unit of classification~\cite{alOmar22refactor}.
Commit messages remain the most frequently used feature for supervised learning, e.g.,~\cite{gharbi19class,zafar19class,alsolai20syslit,levin17boosting,mockus00identifying}.
Provided that the messages are present, complete, and sufficiently accurate, a reliable classifier can be trained.
However, many studies have shown that this is not the case.
Even worse, it is very difficult to identify certain types of refactorings (e.g., \emph{pull up/push down} method) from these messages reliably~\cite{alOmar22refactor}.
Another challenge for obtaining a classifier lies in reliable cross-project and cross-language classification.
While it is quite common to obtain high scores for classifiers trained on single projects using a single programming language, the same task proves to be much more challenging in cross-project/-language scenarios.
All this is the materialization of the second problem that needed to be solved (\textbf{P\=/II}).

In our previous study,~\citepaper{honel2018changeset}, we suggested the source code density as a new, count-based size metric for software.
We hinted that this metric, together with other size-based metrics, might be reasonable predictors for the nature of commits~\cite{honel2018changeset_poster}.
%~\cite{honel2018changeset,honel2018changeset_poster}.
At that time, the state-of-the-art in commit classification achieved an accuracy and Kappa (see~\cite{Cohen1960kappa}) of $\utight{\approx}76$\% and $\utight{\approx}0.63$ in the context of cross-project classification, respectively~\cite{levin17boosting}.
That model is a hybrid one, facilitating features derived from keywords and semantic code changes, of which the latter is rather expensive to compute.
While these performance metrics are impressive, they leave considerable room for improvement.
Hence, analyses that rely on correct classification will inherit this residual error.
Therefore, we conducted a number of consecutive studies to test the aptness of source code density and related size-based metrics for commit classification.

In all of our studies, we made the idealized assumption that any commit can purely be assigned to one of the three maintenance activities \emph{adaptive}, \emph{corrective}, or \emph{perfective}~\cite{swanson76acp}.
While small and tiny commits usually serve a single purpose~\cite{purushothaman05small}, larger commits tend to \emph{tangle} various changes~\cite{kirinuki14hey}.
However, when creating a dataset and manually labeling commits, it became apparent that most large commits can still predominantly be assigned to one principal activity.
However, since most of the commits we observed have at least some impurity, a probabilistic model that assigns discrete classes can never achieve perfect classification accuracy.
Therefore, the models developed by us do predict real-valued class membership.
A maximum a posteriori decision rule then assigns the class with the strongest membership.
A naturally occurring instance of an impure commit with tangled changes is that of a so-called \emph{merge} commit.
These have two or more parent commits.
Rather than serving a purpose of their own, they merely function as a mechanism to combine patches from different branches.
Merging two or more commits of the same maintenance activity would result in a merge commit of the same activity.
However, if the parents have differing maintenance activities, then there is no straightforward way to assign an activity to the merge commit.
Since only a few commits are actual merge commits, we have deliberately excluded them from all our considerations.

%\subsection*{Study:~\citefield{honel19importance}{title}}
\subsection{Leveraging Size-Based Features}\label{ssec:summary-honel19importance}
\paragraph{Summary of~\citepaper{honel19importance}.}
This is our first study to deal with commit classification.
It pursued multiple purposes: (i) to examine statistical properties and probability densities of commit sizes, (ii) to replicate the state-of-the-art models and test the source code density in those, and (iii) to attempt to improve upon the state-of-the-art.

Having the source code density as a new metric at our disposal allows us to challenge existing work.
For example, some authors report strong correlations between the size of a commit and its nature, not considering differences in size resulting from gross vs. net measurements (e.g.,~\cite{hindle08whatdo,purushothaman05small,hattori08nature}).
More specifically, it was reported that large commits are most often of perfective nature and small commits of corrective nature.
Another example is the claim that measuring the size using the number of files or lines of code is irrelevant as either will correlate strongly with the commit nature~\cite{herraiz06comparison}.
Our results demonstrate, however, that this claim does not actually hold, as the evolutionary patterns when using either method of classification are considerably different.
Net notions allow us to observe reduced counts not only for added/modified/deleted lines, but also for files.
For example, if the net amount of changed lines in a file drops to zero, then the number of net-modified files also decreases.
It also enables the investigation of the distribution of net-empty commits.
We learn, for example, that such commits are \emph{never} of type \emph{adaptive}.
This does not come as a surprise, since adding a new feature cannot be done without adding any (net-)code.
Other results, however, can be confirmed.
For example, the smallest commits are most frequently bug fixes, as found previously by~\citeauthor{purushothaman05small}~\cite{purushothaman05small}.

At the time of conducting this study, the state-of-the-art was a hybrid model exploiting commit message keywords and semantic source code changes by~\citeauthor{levin17boosting}~\cite{levin17boosting}.
Features for the source code changes were derived by the ``change distiller'', a tool for extracting semantic changes between two versions of a software~\cite{fluri07changedistill}.
The actual classifier used was a Random forest~\cite{breiman01rf}.
It achieved an accuracy and Kappa of $\utight{\approx}76$\% and $\utight{\approx}0.63$ in the context of cross-project classification, respectively.
We were able to successfully replicate these results using the authors' dataset~\cite{levin17dataset}, within a margin of error (since a Random forest is a type of non-deterministic and probabilistic model).
Having reached this point now allowed us to safely attempt and use size-based features, as well as to train alternative classifiers.

Improving upon the state-of-the-art was approached using a search grid of various classifiers and differently configured datasets.
Similar to~\citeauthor{levin17boosting}'s approach, we first consider various ensemble models that aggregate the results from two underlying models which are trained on different feature sets.
Four different feature sets are used: Keywords, semantic code changes, density and size-related attributes, and a meta feature set combining all these.
This allows us to also compare, for example, how well the source code density functions as a feature when it is used exclusively or replaces another set of features.
The density-only models achieve an average validation accuracy and -Kappa of $\utight{\approx}53.2$\% and $\utight{\approx}0.225$, respectively.
While skillful, those numbers beat the so-called \emph{Zero-Rule} classifier's baseline of $\utight{\approx}43.45$\% only by single digits.
Nevertheless, this implies that the source code density and the other size-based features have at least some predictive power for obtaining a commit's associated maintenance activity.
We conclude that the size of a commit is a significant predictor, with the net size being more important.
A more significant and important result is that the size-based features can serve as a drop-in replacement for the semantic code change features, without a noticeable decline in model performance.
Since we have previously concluded that source code density without the expensive elements of clone detection and string similarities is just as useful~\citepaperp{honel2018changeset}, this replacement considerably reduces the efforts to generate the required features when compared to the semantic code changes of~\citeauthor{fluri07changedistill}~\cite{fluri07changedistill}.

Lastly, we attempt to improve the state-of-the-art.
For that procedure, we essentially ask whether adding the source code density and other size-based features can further boost the classification accuracy and -Kappa.
Our procedure is to first horizontally merge the dataset, then eliminate all gross size, zero- and near-zero variance features.
Then, we run a recursive feature elimination using a Random forest to find the most important predictors~\cite{darst18rfe,kuhn13predmodel}.
Among the ten most important features, we find three density-related attributes, four related to keywords, and three to semantic code changes, meaning that the source code density and size-related features play a meaningful role in predicting the maintenance activity.
In other words, it is quite possible to predict the maintenance activity purely from the metadata associated with a commit.
The champion model found by us leverages all types of features, using a ``LogitBoost'' classifier~\cite{friedman00logitboost}.
In a cross-project scenario, it peaks at $\utight{\approx}89.13$\% accuracy and a Kappa of $\utight{\approx}0.83$, and exhibits a typical accuracy and Kappa of $\utight{\approx}80$\% and $\utight{\approx}0.69$, respectively.
According to~\citeauthor{landis1977application}~\cite{landis1977application}, Kappa values in the range $\interval{0.61}{0.8}$ are considered substantial, and values in the range $\interval{0.81}{1.0}$ are considered as almost perfect.
We also find that contrary to earlier studies, the classification of commits varies across different notions of size.

\subsection{Commit Features of Previous Generations}\label{ssec:summary-honel2020using}
\paragraph{Summary of~\citearticle{honel2020using}.}
This is our second study related to commit classification.
It is an extension to the previous one~\citepaperp{honel19importance}, in that it additionally suggests that information about preceding commits can be used to further improve and stabilize classification performance.
More precisely, in a consecutive chain of commits, this study attempts to predict the nature of the youngest child commit (the so-called \emph{principal} commit) by incorporating features of its direct predecessors.
The approach is based on the conjecture that certain maintenance activities are commonly preceded by certain other maintenance activities, i.e., that commits of a certain type do \emph{not} just appear at random.
In other words, we view a chain of commits as a discrete-time process with not-entirely-random transition probabilities.
The existing dataset by~\citeauthor{levin17dataset}~\cite{levin17dataset} does not contain chains of consecutively labeled commits.
We have reverse-engineered the manual labeling rules (to be found in~\cite{honel2023hmm_dataset}) and asserted their quality using multiple raters (inter-rater reliability).
Then, we created an extended dataset that includes all the commits from~\citeauthor{levin17dataset} and some variable-length chains of consecutive commits.
We attempt various dataset configurations that vary with regard to the features available to the principal commit.
From its predecessors, in any case, only size-based features and keyword counts are used.

We find that it is best to pick a principal commit that uses all available features.
We confirm our previous findings that replacing code-change features with size features does not result in a decline in model performance.
Looking back up to three commits in time improves prediction accuracy considerably.
Models trained for single projects profit even more significantly from considering preceding commits and achieve an accuracy beyond $93$\% with an almost perfect Kappa of $0.88$.
While the previous study~\citepaperp{honel19importance} found that existing state-of-the-art approaches to commit classification can be boosted by $\utight{\approx}13$\% in classification accuracy, this study demonstrated that the source code density of previous commits can further increase that number to $\utight{\approx}17$\%.

\subsection{Relations and Joint Conditional Probabilities}\label{ssec:summary-honel2023hmm}
\paragraph{Summary of~\citepaper{honel2023hmm}.}
Our last study relating to classifying commits into maintenance activities to date is a formalization of the previous extension.
In it, we attempt to exploit relations between adjacent commits, as well as sojourn-times, that is, how much time has passed since the parent commit.
The formalization lies in attempting various types of models that are\,---\,supposedly\,---\,more suitable for this task.
These encompass hidden Markov models, dependent mixture models, and models based on joint conditional densities.
While in the previous extension, we used ordinary classifiers which can only handle \emph{stateless} problems, this study was designed to exploit states and transitions.
The statelessness was previously achieved by horizontally concatenating commits with features of their direct parents.
This study does not attempt to improve the state of the art.
Rather, it seeks to answer the questions (i) whether stateful models can outperform stateless models, (ii) whether simple univariate hidden Markov models are apt for this task, (iii) whether 1st- and 2nd-order dependent mixture models perform better than stateless classifiers, and (iv) whether 1st-, 2nd-, and 3rd-order joint conditional density models can outperform stateless classifiers.

Markovian models, similar to Bayesian networks, are directed acyclic graphs.
In a Bayesian network, we would compute a \emph{conditional} probability by having one or more of its variables assume specific values~\cite{pearl85bayesian}.
Consider the classical \emph{Rain\,---\,Grass wet\,---\,Sprinkler} example.
The \emph{posterior} probability distribution of the grass being wet is conditional on whether it rained, the sprinkler ran, or both.
In the context of commits, we have a discrete posterior probability distribution (for the three maintenance activities) that is thought to be conditional on the attributes of the preceding commits (sometimes also including their labels).
This is called a joint conditional density model, as it is a somewhat generalization of a Bayesian network.
Joint conditional density models are somewhat special, as they are stateless (not requiring features of their parents) while simultaneously having encapsulated state within.
For the task of inferencing, a number of joint conditional density models are tested against a chain of consecutive commits, and the one with the largest a posteriori likelihood is selected (and the label it predicts).

Similar to that, Markov models that make some observations at any discrete point in time, without being able to observe the actual state, are called latent or \emph{hidden} Markov models~\cite{baum66hmm}.
While the state is latent, the transition process is not.
A (discrete) hidden Markov model, therefore, has three additional elements: A vector of initial state probabilities to start in a certain state (out of the finite set of defined states), a transition probability matrix (that is, how likely it is to transition from one state to another), and an emission probability matrix (the probability to observe a certain emission when in one of the states).
When we look at a random commit, for example, it is most likely that it is of \emph{perfective} nature, since perfective commits are the most frequently occurring commits.
If we assume it is indeed perfective, then it is most likely followed by another perfective commit (i.e., we find the highest transition probability from perfective to perfective again).
Since we cannot observe the nature of the next commit but rather only its features, there are certain emission probabilities for the attributes of the next commit.
Since we assume to be at a perfective commit and the next commit being of perfective nature again, the probability to observe new files is likely to be lower, compared to a case where we might expect an adaptive commit next.
A dependent mixture model is a generalization of a hidden Markov model that supports multivariate emissions~\cite{visser10depmix}.

The stateful hidden Markov and dependent mixture models appear to get frequently stuck in what are called \emph{absorbing} states.
Following the earlier example of going from perfective to perfective since it is the most likely transition, it is exactly this scenario these models are vulnerable for.
Univariate hidden Markov models are not able at all to reliably predict the nature of commits (ii).
We observe, however, that the stateless joint conditional density models outperform other stateless, state-of-the-art classifiers, such as Random forest or Gradient Boosting Machine by double digits, both for accuracy and Kappa (iv).
In summary, stateful models are outperformed by simpler, stateless models (i), and this is similarly true for the tested $1$st- and $2$nd-order dependent mixture models (iii).
This was an explorative study, so we have not attempted to produce a usable off-the-shelf solution.
Such a classifier would require a larger training set and more in-depth validation.
Rather, we demonstrated that relational properties and sojourn times among and between commits have predictive power and are worth exploiting.

\cleardoublepage
\chapter{%
Exploiting Software Evolution for Process Quality
%Facilitating Change for Software Process Quality
%%Exploring The
%Aptness of Change for Modeling Software Process Activities
%Aptness of Software Process Activities Modeled Using Change
%Facilitating Software Process Quality For Organizational Learning
%
}\label{chap:obj-2}%
\vspace{5pt}
\epigraph{%
\emph{``Without a repeatable process, the only repeatable results you are likely to produce are errors.''}\;---\;\citeauthor{macala96repeat}~\cite{macala96repeat}%
\\\vspace{20pt}%
\emph{``Since software is the major output of software processes, increasing software process management quality should increase the project's chances of success.''}\;---\;\citeauthor{perkusich15procedure}~\cite{perkusich15procedure}%
}
\restoregeometry
\vspace{5pt}
\chaptertoc
\chaptermark{Software Evolution \& Process Quality}

%\section{Overview and Objective}
%\newpage
\noindent
The contributions of this chapter are mainly based on two publications:~\citepaper{picha_honel22pilot} (refereed) and~\citearticle{honel2023embedded} (revised manuscript submitted for publication), summarized in Sections~\ref{ssec:summary-picha_honel22pilot} and~\ref{ssec:summary-honel2023embedded} through~\ref{ssec:summary-honel2023embedded-end}, respectively.
\textbf{\citepaper{picha_honel22pilot}} is a pilot study aiming to analyze maintenance activities and issue-tracking data from student projects.
The study examines requirements, development, and descoping activities, visualizing cumulative effort and code density.
Experts independently assess the severity of the Fire Drill phenomenon in order to create a ground truth.
The visualizations reveal discernible patterns in projects affected by the Fire Drill anti-pattern.
\textbf{\mbox{\citearticle{honel2023embedded}}} is a follow-up embedded case study in which we revise our approach to modeling activities derived from source code and issue-tracking data.
We measurably improve the reliability of the ground truth by adding another, independent expert.
We use weighted kernel density estimation for issue-tracking activities but found that the source code density had little impact on the probability densities.
We develop a predictive model for detecting and assessing the severity of the Fire Drill.
Using oversampling, we increase the available data and subject it to an adaptive training workflow for robust models.
The workflow demonstrates improved model performance if data is added continuously.
This study emphasizes the need for reliable ground truth and highlights the potential of quantitative insights and variable importance analysis.
It further proposes a replicable framework for organizational learning from quantitative data and highlights the importance of qualitative analysis and structured evaluation for understanding anti-patterns.

\section{Objective and Problems}
This chapter is dedicated to the second objective (see Section~\ref{sec:objectives}).
It summarizes the results of the two publications that we previously recapped.
This chapter's objective and its related problems are defined as follows.

\vspace{10pt}
\begin{mdframed}
    \textbf{O-II}:~\textit{Enable the harnessing of application lifecycle management data for aspects of software process quality and organizational learning thereof.}
    %\textbf{O-II:}~\textit{Examine the suitability of application lifecycle management data, such as version control system data and issue-tracking reports, for the modeling of continuous software processes.}
    %Establish an efficient and effective size-based change detection that can be used for capturing the integral aspects of software evolutionary processes.}
\end{mdframed}
\vspace{5pt}

\noindent In order to achieve the objective, the following problems need to be solved:

\begin{enumerate}[itemsep=.5ex, leftmargin=3.5em]
    \item[\textbf{P\=/III:}] Identify whether, which, and how application lifecycle management data can be leveraged for process modeling.
    %\item[\textbf{P\=/IV}:] Determine whether detecting process-related problems that originate in management are reflected in such data.
    \item[\textbf{P\=/IV}:] Obtain a stable predictive model for assessing anti-pattern presence and -severity.
    \item[\textbf{P\=/V}:] Find a replicable methodological framework that enables automatic organizational learning from a few past software projects.
    %\item[\textbf{P\=/III:}] Suggest a common and adequate representation of the data for continuous processes.
    %\item[\textbf{P\=/II:}] Obtain a reliable and accurate change classifier that generalizes beyond single projects and/or languages.
\end{enumerate}

\section{%
Activity-Based Process Modeling%
%Modeling Of Processes%
%Process Modeling Using Activities%
}
%
%We need a common, mathematical representation of the process, such as dynamic process.
%Dynamic means changing over time.
The quality of a software product is regularly governed by software quality models.
Such models are used to monitor and eventually achieve a previously defined quality goal~\cite{Singh2013reviewqms}.
Most commonly, software quality models facilitate \emph{attributes data}, that is, metrics that describe the current condition of the software product~\cite{carleton1999measure}.
Therefore, these models can capture and describe merely a snapshot of the product they quantify.
While we could, for example, compare consecutive snapshots to follow the evolution of the product's quality, the current models themselves cannot encompass or portray the dynamics involved in a software process.
Obtaining and evaluating software metrics continuously has been done previously.
For example, we might measure \emph{Code Coverage} or \emph{Cyclomatic Complexity} on an ongoing basis~\cite{kitchenham1995,McCabe76}.
Doing so is perhaps more relevant in cases where the metrics' absolute values are less meaningful than their change over time.
Rather, to derive a notion of quality, the interest lies in \emph{how} the change $\Delta_{\mathsf{m}}$ of some metric \textsf{M} came to be.
%
% Later: Make point with the Fire drill which is sensitive to a balance of activities; it does not care so much about amounts
%
Existing quality models concern the software product much more than the software process.

We have previously introduced the new software metric of source code density (see Chapter~\ref{chap:obj-1}).
While useful as a metric, its greater utility lies in the ability to confidently predict the maintenance activity associated with a commit~\cite{swanson76acp,honel2020using}.
The ability to capture the carried-out activities during the software process allows for defining related quality goals which are tied to expectations of how each activity may unfold.
For example, we might expect a certain amount or trend of an activity over a defined time frame.
Another example would be an expectation that relates to the combined behavior of two or more activities.
Therefore, we suggest modeling expectations and observed instances of activities as continuous-time dynamic processes.
This approach also has the advantage that it preserves a human-understandable representation, which~\citeauthor{curtis92process} defined as the first goal of process modeling~\cite{curtis92process}.
A similar approach of modeling the temporal accumulation of activities was previously applied in the so-called Unified Process~\cite{Scott2001uniproc}.
The Unified Process, however, calls these \emph{disciplines} (e.g., business modeling, implementation, test, etc.) and horizontally segments them into discrete phases, such as inception, elaboration, or construction.

\citeauthor{swanson76acp}'s maintenance activities (\emph{adaptive}, \emph{corrective}, and \emph{perfective}) provide a great insight into the developers' day-to-day software engineering activities~\cite{swanson76acp}.
Unfortunately, they cannot reflect all the various disciplines as modeled in, e.g., flavors of the Unified Process.
While source code and commits thereof are deemed to be an objective data source, their aptness for predicting the associated maintenance activity is subject to a classifier with high confidence.
In other words, the residual error in the classifier will be reflected in subsequent models that operationalize maintenance activities.
Worse, there does not exist a reliable way to precisely estimate the effort (in terms of time spent) that went into producing a commit~\citepaperp{honel2018changeset}.
However, software repositories are just one type of artifact found in application lifecycle management data.
Another, even more, direct reflection of the time spent on certain activities is to be found in the \emph{issue-tracking} data~\cite{chappell2008application}.
Issue reports, however, are no panacea, either, as they are subject to inaccuracies, such as misclassification of tickets, or misestimation of the time spent.
Issue-tracking data tend to be comparatively more scarce, as it receives fewer updates.
Often, items that are deemed too small are never logged.
We suggest that both of these application lifecycle management data sources, source code and issue-tracking, allow to (at least partially) portray some of the most vital disciplines that are required for the detection of a class of software process problems adequately.

\section{%
Problem Detection%
%Detection Of Problems% in the Software Process
}
Detecting a problem (a quality regression) in the software product can be done effectively using various software quality models.
For that, a number of approaches and standards, such as ISO/IEC 25010, exist.
As software is the major output of software processes, a causal relation exists~\cite{perkusich15procedure}.
\citeauthor{halvorsen2001} directly suggest the relation $\mathrm{Quality}\!\left(\mathit{Process}\right)\Rightarrow\mathrm{Quality}\!\left(\mathit{Product}\right)$~\cite{halvorsen2001}.
\citeauthor{kruchten2003rup} enumerates typical root causes and symptoms encountered in software processes, such as an inaccurate understanding of end-user needs, the inability to deal with changing requirements, the late discovery of serious projects flaws, team members obstructing each other, or the lack of trust~\cite{kruchten2003rup}.
This implies that many types of problems have their origin in bad practices of the developers and/or project managers.
A (good) practice is ``a proven way or strategy of doing work to achieve a goal that has a positive impact on work product or process quality.''~\cite{omg2008spem}
Still, approximately half of all IT projects fail because of bad practices~\cite{emam08failures}.
Therefore, attempting to study problems and learning from past processes is a valuable opportunity for organizational learning~\cite{neill2005antipatterns}.
There exists a body of work concerned with detecting problems in software processes.
A widely applied approach is that of simulation modeling, which appears to be implemented most often using system dynamics, discrete events simulation, and Bayesian networks~\cite{kellner1999procsim,perkusich15procedure}.
In a systematic review, \citeauthor{zhang2008procsim} unveil the other most frequently used implementations, none of which currently exploits our suggested approach of continuous-time dynamic processes~\cite{zhang2008procsim}.

\subsection{%
%Patterns And
Anti-Patterns%
}
The definition of what constitutes quality does not necessarily have to follow a standard, as the meaning of quality can be subjective and application-specific~\cite{prahalad1999quality}.
Observing activities as the underlying quantity for evaluating the quality of a software process grants access to defining and monitoring a certain domain of quality goals.
More specifically, there exist a large number of \emph{anti-patterns} in the context of software engineering~\cite{brada2019catalogue}.
Many such anti-patterns concern bad coding practices such as tight coupling, lack of abstraction, or duplicated code and can be detected quite well using, e.g., static code analysis.
However, another domain of \emph{project management} anti-patterns, which are concerned with bad practices in the form of activities as carried out by the developers or management, exists~\cite{brown1998antipatterns,brown2000anti,neill2005antipatterns,stamelos2010software,brada2019catalogue}.
We argue that these kinds of bad practices are particularly well suited to be modeled using activities as can be found in the application lifecycle management data.
Previous approaches of limited utility that used activities for detecting exist~\cite{picha2017activity,picha2019apdetect}.
However, we suggest not only detecting but accurately assessing the severity of the manifestation of anti-patterns.

A pattern is a reusable solution to a problem that has been identified in the past, while an anti-pattern is a repeated behavior or design choice that has been found to be detrimental to an application's quality. %Patterns can help us make better decisions when faced with similar problems, while anti-patterns should be avoided as much as possible since they often lead to poor outcomes.
Patterns are complex phenomena that are most often described using a ``pattern language''.
They offer a solution to a class of problems that is intentionally left vague~\cite{brown2000anti,neill2005antipatterns}.
Patterns are especially prevalent in phases of design, project management, and in software development processes~\cite{rising2000scrum}.
%A pattern provides a general and proven solution to a common problem.

\begin{displayquote}
    \textit{``Each pattern describes a problem which occurs over and over again in our environment, and then describes the core of the solution to that problem, in such a way that you can use this solution a million times over, without ever doing it the same way twice.''}\vspace{-15.5pt}\begin{flushright}
    ---\,\citeauthor{alexander1977pattern}~\cite{alexander1977pattern}
    \end{flushright}
\end{displayquote}

An anti-pattern is some kind of problematic software process situation and is often the result of malpractice or human error.
As a result, anti-pattern project management phenomena pose threats to project quality and -delivery.
The outcomes include but are not limited to, developer churn, interpersonal and organizational tensions, a product of poor quality, delayed delivery, or even total project failure.
%which often result in, e.g., a final product of poor quality or increased tensions within the organization.
The notion of an anti-pattern, therefore, is closely related to that of \emph{project risk}~\cite{stamelos2010software}.

\begin{displayquote}
    \emph{``While it is certainly useful to study the successful ways people solve problems, the old adage that we learn from our mistakes suggests that studying failures might be even more fruitful.''}\vspace{-12.5pt}\begin{flushright}
    ---\;\citeauthor{neill2005antipatterns}~\cite{neill2005antipatterns}
    \end{flushright}
\end{displayquote}

A prominent anti-pattern is the so-called \textbf{Fire Drill}~\cite{silva2015spm,brown2000anti}.
It is characterized by management forcing the development team to deliver immediately after a longer slack time~\cite{nizam2022failure}.
As per the Fire Drill's description, longer delays are usually caused by management.
The reasons for that are manifold: project uncertainty, poor communication, a dysfunctional organization, or poor resource allocation are just a few of the possible and often intertwined reasons.
The Fire Drill caught our focus because it affects open source and industrial projects alike.
We chose to analyze it as our assumption was that even though it is a managerial anti-pattern, it would be sufficiently well reflected in the activities as derived from the application lifecycle management data.

\subsection{Operationalization of Anti-patterns}
The operationalization of Anti-patterns is difficult, due to a lack of formal descriptions and models thereof.
Most predominantly, they are described using a pattern language, without a concrete solution, but rather a prototype for it~\cite{neill2005antipatterns}.
However, there have been a few approaches to operationalizing anti-patterns by representing them using a more actionable form.
For example, Bayesian (Belief) networks, ontologies, social networks, and Design structure matrices have been used previously to represent aspects of anti-patterns~\cite{settas2006using,settas2007using,stamelos2010software,perkusich15procedure}.
Such methods, however, are mostly concerned with encoding project management knowledge into a machine-readable and actionable form.
None of the methods allows to detect the presence of a certain anti-pattern.
Most models are concerned with a specific aspect, such as assessing uncertainty or exploring relations~\cite{fenton2004making}.

\section{Aptness of Activities for Process Modeling}
While activities were previously accumulated in so-called \emph{disciplines} in flavors of the Unified Process~\cite{Scott2001uniproc}, no practical attempt for the operationalization of those for the presence detection of anti-patterns had been made.
Furthermore, we suggest to use maintenance activities as predicted from source code commits and reports from issue-tracking software to derive activities such as development, forward engineering, requirements elicitation, etc.
Those activities only partly reflect the disciplines of the Unified Process.
While there is an overlap, some activities, such as \emph{descoping}, are not found as disciplines, and some disciplines have no (full) counterpart in the activities.
Therefore, addressing problem \textbf{P\=/III} needed to be approached exploratively.

\subsection{Exploring Activities}\label{ssec:summary-picha_honel22pilot}
\paragraph{Summary of~\citepaper{picha_honel22pilot}.}
We designed a pilot study that would visualize the maintenance activities (\emph{adaptive}, \emph{corrective}, and \emph{perfective}) from source code, as well as three activities from issue-tracking data.
These are \emph{requirements} (cumulative effort spent in the project\,---\,``hours burnup''\,---\,on requirements, analysis, and planning), \emph{development} (implementation, testing, and fixing of bugs), and \emph{descoping} (development effort that was planned but never spent).
For that, we examine the application lifecycle management data of $15$ student projects~\cite{honel2023fddataset}.
Over the course of several years, students would participate in a software engineering course to learn agile and iterative practices.
In groups of four students (average), they would design a software product over the course of approx. six iterations, corresponding to a duration of ca. three months.
The customer for each project usually comes from the industry, such that the groups develop a real-world product with actual requirements.

On average, a project has about $102$ issues and ca. $158$ commits.
In some cases, even fewer issues were reported.
In other cases, much time would pass between issues, bringing the logging to a halt.
Therefore, we decided to model issue-tracking activities cumulatively, adding up the time that was spent on each activity and then normalizing this quantity such that it would have a range of $\interval{0}{1}$.
For source code, we decided to model the temporal accumulation of the three maintenance activities using kernel density estimation~\cite{rosen1956kde,parzen1962kde}.
This was done because no reliable method of estimating the effort that is associated with each commit was available.
We would also model the source code density itself as a linearly interpolated function.

Prior and independently to visualizing all of the activities, two experts would do a qualitative analysis of whether the Fire Drill was present in each of the projects.
The experts would make a numeric rating, assessing the severity on a numeric rating scale of $\mathbf{0\tight{-}10}$, where $0$ indicates the absence of the phenomenon and $10$ is the strongest possible manifestation.
The visualization of all activities across the projects made clear that\,---\,in a subjective way\,---\,the activities of affected projects would manifest with a different and discernible pattern.

\section{A Framework for Activity-Based Phenomena}\label{ssec:summary-honel2023embedded}
\paragraph{Summary of~\citearticle{honel2023embedded}.}
The remainder of this chapter is dedicated to summarizing the changes between~\citepaper{picha_honel22pilot} and~\citearticle{honel2023embedded}, as well as the extensions, framework, and results contributed by the latter study.

\subsection{Consistent Activity Design}
In the follow-up embedded case study (\citearticle{honel2023embedded}), we revised which and how we would model activities derived from source code and issue-tracking data, based on our previous findings.
The main reason was to represent activities using a single, uniform way.
For issue-tracking activities, we used weighted kernel density estimation, facilitating the duration of instances of each activity.
For source code, however, there is no precise or reliable notion of how much time was spent on a commit~\citepaperp{honel2018changeset}.
Hypothetically, the source code density should correlate somewhat with the \emph{effort} of producing a commit, which could be used as weights.
However, the source code density in the projects has an expectation and standard deviation of $\utight{\approx}0.799$ and $\utight{\approx}0.19$, respectively~\cite{honel2023dissrepl}.
When modeling the probability densities for each of the activities, using the source code density as an additional weight would not alter the probability densities significantly, but only add tiny perturbations to it.
Because of this and our results from earlier that showed, while significant, only a poor correlation between source code density and effort, we decided not to use the source code density as an additional weight during kernel density estimation.

\section{Prediction of Presence and Severity}\label{sec:obj2-predict}
The next and perhaps most practically relevant problem, \textbf{P-IV}, was to find a predictive model that would exploit the activities and detect the presence of and/or assess the severity of the Fire Drill manifestation.

\subsection{Expert-engineered Rules}
The cumulative and normalized design of activities was chosen in our pilot study for another reason.
Prior to examining any project-related artifacts, experts would discuss and agree on expected thresholds for each of the activities from which indicators could be derived.
Multiple indicators could then be combined to define various binary decision rules, that would either indicate the absence or presence of the anti-pattern.
For binary classification, this rule achieved significant accuracy.
However, accuracy is not a \emph{proper} scoring rule to assess the performance of a decision rule when binary outcomes are to be compared to a numeric linear rating, which can be scaled to probabilities~\cite{gneiting2007proper}.
Therefore, the performance was measured more adequately using~\citeauthor{brier1950verification}'s score, since it is a strictly proper scoring rule~\cite{brier1950verification}.
The score, which is technically equivalent to the mean squared error, expresses the deviation between the classification result (either zero or one) and the class probability continuously, where $0.0$ would indicate a perfect score and $1.0$ a maximal deviation.
The score would then reveal that the decision rule, probabilistically, performs worse than the so-called Zero-rule, which is the least skillful probabilistic model that constantly predicts the expectation of the dependent variable, here $0.5$~\cite[p.~183]{honel2023fdtr}.

For maintenance activities derived from source code, no experts existed, as using such data for this task is novel.
We concluded that expert-engineered rules are not suitable for detection.
This is especially true for binary decision rules, as a complex phenomenon such as the Fire Drill has many facets to its possible manifestations.
For a proper severity assessment, we would need to engineer features and fit a regression model, especially for the source code data.

\begin{figure}[t!]
    \centering
    \makebox[\textwidth][c]{\input{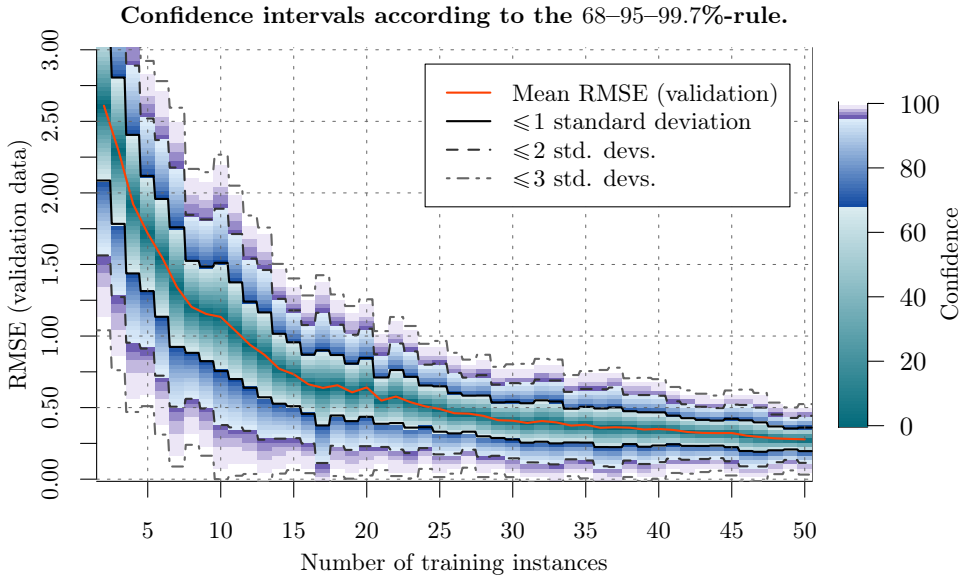}}
    \caption[Adaptive training workflow with confidence intervals, per number of projects.]{{\small Continuous confidence of the neural network predictor, with regard to the number of training instances. Shown are the values according to the $68$--$95$--$99.7$\%-rule (assuming a normal distribution for every generalization error). The three color gradients correspond to the three sigmas.}}
    \label{fig:rr-3sigma-robust}
\end{figure}

\subsection{Effective Training on Scarce Data}
Training a predictive model in scenarios similar to ours, that is, having only a few projects available, is challenging.
In fact, the number of features greatly outnumbers the projects at our disposal.
Therefore, the synthetic minority oversampling technique was applied in order to increase the amount of available data~\cite{torgo13smote}.
We then designed a training workflow that would follow recommendations for obtaining robust predictive models under constraints of small sample sizes (e.g.,~\cite{cawley2010overfitting,raudys91small,vabalas2019}).
Another limitation that was encountered in the preceding pilot study was uncertainty about the quality of the two raters' assessments.
Therefore, we added a third independent rater to ensure a minimum quality, as the training of the regression models would strongly depend on it.
The rigorous adaptive training workflow that was evaluated over a very large grid using many repeats demonstrated that successively adding training data continuously improves model performance.
This means that the source code and issue-tracking activities and engineered features thereof are indeed suitable to predict complex phenomena such as the Fire Drill.

Learning from between $\mathbf{12\tight{-}20}$ instances (projects) will yield a regression model of practical useful precision.
Perhaps most notably, training the champion model (a simple neural network) on $25$ instances produces an expected validation error of $\utight{\approx}0.46$, and this error will not be larger than $\utight{\approx}0.96$ by a probability of $99.7$\% (three standard deviations) on the severity scale of $0\tight{-}10$.
In other words, predicting the severity of the manifestation in a project will almost surely be off by less than one, on a scale of zero to ten.
This and other results are shown in Figure~\ref{fig:rr-3sigma-robust}.
The mean \emph{Root Mean Squared Error} (RMSE) was determined using predictions on validation data that was not part of the training.
Therefore, it allows us to approximate the empirical generalization error.
The three color gradients correspond to the three sigmas (standard deviations from the mean).
Figure~\ref{fig:rr-3sigma-robust} portrays the adaptive training of the champion model only.
Estimates for the expected validation error were obtained using the so-called \emph{Leave-One-Out Cross-Validation} (LOOCV) empirical risk estimator~\cite{lachenbruch1968Loocv}.
Each training was repeated $50$ times, using a nested grid search.
\citeauthor{Varma2006cv}~\cite{Varma2006cv} and~\citeauthor{vabalas2019}~\cite{vabalas2019} have previously demonstrated that standard K-fold cross-validation produces strongly biased performance estimates, particularly with small sample sizes.
However, measures to avoid this problem include, for example, using some form of nested cross-validation, train/test split approaches, and sufficiently many repeats, all of which were implemented~\cite{shaikhina2015small}.

The residual errors were almost always distributed normally.
This allowed us to apply the $68$--$95$--$99.7$\%-rule (often called the $3\sigma$-rule) for approximating a confidence interval, per number of training instances.
The figure shows that the expected generalization error decreases quickly when we keep adding projects to the training data in the beginning.
Generally, the adaptive training process appears to converge rather well with exceeding amounts of training data.
However, the results for the training using $\utight{\geq}15$ projects should be treated with caution, as additional data was obtained using the \emph{Synthetic Minority Oversampling TEchnique} (SMOTE) for regression.
Yet, SMOTE has previously shown the ability to significantly improve model performance~\cite{torgo13smote}.
Therefore, the results obtained by training with $\utight{\geq}15$ projects might still give us a somewhat valid indication as to the expected generalization error.

\section{Efficient Organizational Learning}\label{ssec:summary-honel2023embedded-end}
Organizations frequently embrace failed projects as valuable opportunities for organizational learning~\cite{birk2002postmortem}.
Using anti-patterns for \emph{efficient} organizational learning from past projects is inhibited by a variety of factors today.
While thorough qualitative analysis as conducted during, e.g., project post mortems is a widely accepted practice, it requires experts that are knowledgeable and available in the problem domain~\cite{Matsubara2021}.
Qualitative analysis usually is labor-intensive and prone to errors, as human experts tend to introduce their own subjective bias~\cite{muram2018}.
The application lifecycle management that governs software processes provides a multitude of digital artifacts, most of which cannot currently be leveraged for anti-pattern instance and -severity detection~\cite{picha2017towards}.
A strong inhibitor for organizational learning is perhaps the lack of a quantitative description for anti-patterns~\cite{Simeckova2020spem}.
The scarcity of available historical data in the shape of past projects increases the difficulty of properly abstracting and generalizing from the lessons learned.

\subsection{Quantitative Data}\label{ssec:obj2-quant-data}
Efficient organizational learning from a few projects which are affected by anti-patterns, that exploits activities as derived from application lifecycle management data, requires a \emph{meta-process}.
In our embedded case study of the Fire Drill, we suggest a replicable methodological framework for this task.
Specifically, the framework enables automatic learning from quantitative data.
It requires experts only initially to produce a \emph{reliable} and reusable ground truth.
In our case, three experts were sufficient to achieve a mostly substantial agreement on the Fire Drill's severity in the examined projects~\cite{landis1977application}.
If the ground truth converges with increasing numbers of experts and amount of knowledge towards a desired quality, then it can be facilitated for exploiting the mostly quantitative artifacts and the process of adding can be stopped.
Convergence can be measured by calculating the inter-rater agreement and benchmarking the result~\cite{gwet2008ac1}.

We were able to confirm our hypothesis that the Fire Drill is sensitive to a certain balance of activities.
This anti-pattern shares similarities (in terms of indicators, symptoms, or consequences) with many other anti-patterns, such as ``Collective Procrastination''~\cite{picha2019apdetect}, ``Brook's Law''~\cite{brooks1995myth}, or ``Half Done Is Enough''.
These can likely be modeled using activities, too.
Therefore, the framework is likely to be applicable to such related phenomena.
We demonstrate how the Fire Drill manifests in the modeled activities from source code and issue-tracking, by visualizing the activities as weighted mixtures, which are convex combinations.
In a convex combination, the probability density for a single activity across all projects is combined using the weight of each project, which was derived from the experts-mined ground truth.
Those mixtures exhibited a characteristically strong behavior that was in concordance with the qualitative findings from the case study and helped us to understand deviations from those findings, too.
The weighted mixtures, therefore, represent how the Fire Drill typically unfolds and manifests in the projects affected by it.
If more and more projects are added in the future, the mixtures may even become a type of \emph{quantitative} pattern description that could be used in, e.g., predictive models.

Another important quantitative insight can be derived from variable importance.
It is predominantly used to select the features that should participate in a predictive model~\cite{zhu2015varimp}.
We demonstrate how it can be leveraged for organizational learning from a quantitative perspective.
Of course, the types of insights to be gained depend strongly on the type of features that were engineered.
However, a common feature in our presented framework for the detection and severity assessment for activity-based anti-patterns is that of segments, that is, the subdivision of activities along the time axis.
The importance of each segment allows us to learn about, e.g., the criticality of the examined phenomenon in certain project phases.
Furthermore, those segments may be aligned more deliberately to overlap with phases of the governing process model (e.g., ``construction'' in the Unified Process~\cite{Scott2001uniproc}).
Two more types of features came naturally: the amount of each activity and the difference (computed as a divergence) between any two activities in each segment.
These two universal features taught us, for example, that the amount of any carried-out activity is less important than the balance between activities.
In other words, the presence and severity of a Fire Drill can be determined more accurately just by knowing how imbalanced two activities were, rather than knowing about the absolute amount that was carried out in a segment.
We have not yet explored other, obvious possibilities for types of features, such as trends, which might be good choices for this and other phenomena.

Lastly, the adaptive training workflow was designed to accommodate any arbitrary set of features and any number of data points.
In the case of convergence, as a byproduct of obtaining a skillful predictive model, we also learn about which type of model works best (e.g., neural network, random forest, etc.) and how much training data is required.
More specifically, we can obtain precise confidence intervals and expected validation errors and how they correspond with each project added to the training data.
The adaptive training was specifically designed to counter and overcome the negative impacts that data scarcity and over-fitting have~\cite{cawley2010overfitting,vabalas2019}.
The convergence alone indicates that the digital artifacts, activities, and features thereof are, in fact, suitable for the task of anti-pattern detection.
The framework was designed such that it can be replicated in similar settings (e.g., industrial, different phenomenon, etc.).
If the adaptive training would not converge in other cases, then this could mean that the sought-after phenomenon is not well represented using the selected activities.
% be another valuable insight.

\subsection{Qualitative Data}
Organizational learning is commonly achieved by evaluating the often unstructured artifacts and data through experts~\cite{muram2018}.
For anti-patterns, expert-based evaluation still is the current go-to methodology~\cite{picha2019apdetect}.
While we have previously shown how to leverage quantitative data for organizational learning, it is often the qualitative data that provides the background for the hows and whys.
The framework for organizational learning that we introduced previously rests, in fact, on qualitative evaluation, as it requires experts to find a consensus on the examined problem's facets first.
To support qualitative evaluation, certain measures to improve its objectivity and robustness can be implemented.
A plethora of guidelines for systematic evaluation exists, especially in the context of case studies (e.g., \citeauthor{yin2013}~\cite{yin2013}).
In the context of empirical software engineering, dedicated guidelines can be followed to further structure the analysis~\cite{runeson2009guidelines}.
This approach is somewhat similar to conducting a systematic literature review, where the strong emphasis is on the repeatability of the process.
Strengthening qualitative analysis and increasing the robustness of results can be achieved through observer- and data-triangulation~\cite{runeson2012}.
The agreement between observers can be measured and benchmarked using inter-rater agreement (e.g.,~\cite{gwet2008ac1,landis1977application}).
In our most recent embedded case study, we implemented a process that follows these guidelines and is replicable in similar contexts.
It was specifically designed to support subsequent studies of phenomena similar to the Fire Drill, which are always embedded in some context (e.g., academic, agile, industrial, waterfall, etc.) and, therefore, require thorough analysis.

Qualitative analysis of phenomena such as anti-patterns can potentially grant insights into the studied phenomenon's prevalence and nature (how it manifests).
It may also allow gathering evidence for the \emph{absence} of the problem, as was the case in our embedded case study.
Patterns come only with fuzzily described solutions, which leaves room for instance-specific measures.
In the case of the Fire Drill anti-pattern, this also largely applies to its described symptoms and consequences.
Only qualitative analysis can identify and ascribe problem instances to such loosely described symptoms and consequences.
During our analysis, new supercategories for symptoms and consequences emerged, as those were more explicit to collect instances of observed problems.
For example, the most often reoccurring problems in our case were high project risk, poor communication, and a compromised project schedule or -scope.
We also learned that\,---\,within the confines of the studied context\,---\,not all of the theoretically available \emph{refactored solutions} of the Fire Drill can be implemented.
It is those kinds of results that only qualitative evaluation can reveal.

\cleardoublepage
\chapter{%
%Exploiting and
%Operationalization of Activities for Organizational Learning%
Towards Accessible Organizational Learning%
}\label{chap:obj-3}%
\vspace{5pt}
\epigraph{%
\emph{``Bluntly, the code metric values, when inspected out of context, mean nothing.''}\;---\;\citeauthor{gil2016meannothing}~\cite{gil2016meannothing}%
}
\restoregeometry
\vspace{5pt}
\chaptertoc
\chaptermark{Accessible Organizational Learning}

%\section{Overview and Objective}
\noindent
The contributions of the first half of this chapter are mainly based on the two publications~\citepaper{honel2022qrsmas} (refereed) and~\citearticle{honel2023masjoss} (manuscript submitted for publication), summarized in Sections~\ref{ssec:summary-honel2022qrsmas} and~\ref{ssec:summary-honel2023masjoss}, respectively.
In the second half, starting with Section~\ref{sec:ac}, we contemplate an approach that extends the methodology and results of these publications.

In \textbf{\citepaper{honel2022qrsmas}}, we argue that raw software metrics lack meaningful comparisons or aggregations due to, e.g., different scales and ranges.
A contextual approach assigns scores by leveraging ideal values and observed data, transforming metrics into having a uniform distribution afterward.
This transformation applies to various numeric quantities, including software metrics.
Metrics vary significantly across application domains, requiring careful consideration for accurate quality determination.
Validation against domain-specific quality models is crucial to avoid misleading results.
In \mbox{\textbf{\citearticle{honel2023masjoss}}}, we introduce a tool- and (visual) analysis suite that implements said contextual approach.

\section{Objective and Problems}
%\vspace{10pt}
%\noindent
This chapter is dedicated to the last of the three objectives (see Section~\ref{sec:objectives}).
In the first half, it summarizes the results of the two publications that we previously recapped.
In the second half, we contemplate a potential solution to the last problem.
This chapter's objective and its related problems are defined as follows.

\vspace{10pt}
\begin{mdframed}
    \textbf{O-III:}~\textit{Enhance the explainability of quantitative data and reduce the amount of required qualitative analysis.}
\end{mdframed}
\vspace{5pt}

\noindent In order to achieve the objective, the following problems need to be solved:

\begin{enumerate}[itemsep=.5ex, leftmargin=3.5em]
    \item[\textbf{P\=/VI:}] Find a technique for raw quantities that allows them to be understood, compared, and associated with quality.
    \item[\textbf{P\=/VII:}] Suggest a solution that can provide details about phenomenon severity and serve as a data-based drop-in replacement for qualitative evaluation.
\end{enumerate}

\section{Black- and White-Box Models}
Models considered to be ``white-box'' models use a reasonably restricted number of components and provide traceability and transparency in their decision-making~\cite{Rai2019xai}.
The previous chapter concluded with the results of our most recent case study~\citearticlep{honel2023embedded}, one of which is a predictive model for the Fire Drill severity.
In contrast to white-box models such as Bayesian networks, additive models, or linear regressions, this model is considered a ``black-box'' model.
We can inspect the architecture and weights of such a model, but the degree of mechanistic insight into such a system is low as it defies human comprehension, most often because of non-linear mechanics, too-large capacities, or both.
We identify two problems that prominently inhibit the understanding of black-box models, which are the incomprehensibility of the inputs, and  too-complex aggregations thereof.
The result of these circumstances is a lack of a clear association between the data passed into a model and the qualitative meaning of the result.
For example, if our champion model for predicting the Fire Drill were to output an unexpected severity, then perhaps only a thorough qualitative analysis would allow us to find out why.

The definition of what constitutes \emph{understandability} is subjective and often context-dependent~\cite{zagzebski2017whatis}.
Consider a predictive model that uses the metrics \emph{Lines Of Code} and \emph{Cyclomatic Complexity} as its inputs.
Only the latter of both metrics has a meaning, that is, lower complexity is more desirable.
Yet, while we know that the lowest possible value for it $\utight{=}1$, achieving this value is unrealistic for a reasonably sized software~\cite{McCabe76}.
Worse, \emph{Cyclomatic Complexity} is unbounded, i.e., it lies in the interval $\interval{1}{\infty}$.
Lastly, even if we had knowledge about the practical range of values this metric typically assumes, how can we know that any two unit changes can be considered equally good/bad?
For example, is a decrease in complexity from ten to nine equally valuable as a decrease from two to one?
As for \emph{Lines Of Code}, the problem is further exacerbated, since this metric does not even have any such meaning as, e.g., lower is better.
What constitutes a desirable value is highly context-sensitive.
In fact, most software metrics are meaningless when captured out of context~\cite{gil2016meannothing}.
These circumstances inhibit any two quantities (such as software metrics) to be properly understood, compared, or to be associated with a notion of quality.
In Section~\ref{sec:transf-to-scores} we offer a solution to these problems, by transforming quantities using a previously derived ideal value into \textbf{scores}.
Any two scores possess the same traits: they are scaled into the range $\interval{0}{1}$, higher is better, and all unit increments are considered to be equally valuable, making them perfectly comparable.
Consequently, we characterize understandability as transforming quantities into their respective scores.
A characterization of added understandability in the context of predictive (black-box) models is to exploit scores in conjunction with variable importance~\cite{zhu2015varimp} in order to replace certain aspects of organizational learning that would otherwise have to rely on qualitative evaluation.
In the context of phenomena that unfold over time (as are activity-based anti-patterns), using scores as inputs and simultaneously knowing about their importance can be leveraged to assess single segments in which the severity is high or low.

The champion model used in the embedded case study for Fire Drill severity prediction is a typical black-box model:
we do not understand its inputs, and we most likely cannot comprehend its inner workings, since it is a multilayer perceptron (computations of computations implemented as a feedforward artificial neural network) with non-linear activations.
The features that served as inputs were z-standardized and in some cases pre-processed using principal components analysis~\cite{pearson1901pca}.
Z-standardization scales and translates all quantities to have zero mean and unit variance, which is detrimental to the interpretability of the original quantity.
Decomposing features into their principal components is a ``one-way street'' for interpretability, too, as the components cannot be clearly associated with or traced back to the original features afterward.
Yet, this model is able to predict the desired outcome with an acceptable error, thereby allowing us to bypass the expert-based and expensive qualitative evaluation that would be otherwise required to get to a similar result.
Thanks to the extensive evaluation of that model, we can be reasonably certain that the prediction is accurate.
The question, however, is how that model derives its answer from the inputs we provided.
Suppose that, for a previously unseen project, that model predicts a severity of six out of ten.
If we would want to know \emph{how} this severity came to be, the only option would be to conduct a qualitative analysis of the artifacts, which\,---\,compared to the model\,---\,requires experts, is expensive, labor-intensive, and error-prone and often requires the evaluation of large bodies of unstructured information~\cite{Matsubara2021,muram2018,samalikova2014}.
This circumstance is reflected in problem \textbf{P\=/VII}.

\section{Transformation to Scores}\label{sec:transf-to-scores}
Problem \textbf{P-VI} is perhaps more intricate than it appears at first.
We have previously given the obvious example of standardized or pre-processed features that aggravate their interpretation.
Consider the following example of the software metric \emph{Number Of Packages} with a value of $\mathsf{NOP}\tight{=}23$.
Suppose the quality model is not abstract, i.e., there is a mathematical definition (an aggregation scheme) for how this and other metrics relate to a defined quality goal, such as \emph{maintainability}~\cite{wagner2015quamoco,bouwers2013software}.
If this metric is used as input to a predictive model, such as a software quality model, we would know what this value means and could make sense of it.
However, we cannot know whether this value is a ``good'' value, that is, without \emph{context}, it is impossible to decide whether (and to what degree) $23$ packages are appropriate for the software in question.
The context may be given by, e.g., the software's corresponding application domain.

\subsection{Metrics as Scores}\label{ssec:summary-honel2022qrsmas}
\paragraph{Summary of~\citepaper{honel2022qrsmas}.}
Representing a raw software metric as a score attaches a meaning such as good/\allowbreak
bad, acceptable/\allowbreak
alarming, or common/\allowbreak
extreme to it.
Raw software metrics cannot generally be compared or aggregated because they have different scales and ranges~\cite{ulan2018qms}.
However, being able to compare software metrics, especially across projects or application domains is vital, since it was previously shown that such metrics are meaningless when captured out of context~\cite{gil2016meannothing}.
In a recent study, we have shown that freestanding software metrics cannot be compared to other metrics and are of limited use because there is no quality directly associated with them.
However, we introduce an approach that leverages contextual information and ideal values that eventually allow us to assign a score to a concrete metric value.
The previous example of $\mathsf{NOP}\tight{=}23$ can now be properly solved.
The transformation to a score is done using a context first, which, for software applications, can be given by the application domain (e.g., database application, game, middleware, etc.).
Then, for each context, there is an assumed \emph{ideal} value, which may be user-defined or derived using, e.g., non-parametric methods, such as the expectation $\mathbb{E}\left[\mathcal{X}\right]\tight{=}\int_{-\infty}^{\infty}\,x\,f_{\mathcal{X}}(x)\,dx$, the mode $\hat{x}\tight{=}\mathrm{arg\,\!max}_{x\tight{\in}\mathcal{X}}\,f_{\mathcal{X}}(x)$, the median ($50$th percentile), and observed infimum/supremum (where $f_{\mathcal{X}}$ is the probability density of the random variable $\mathcal{X}$).
The ideal value, together with context-typical observed values, is then used to transform the raw software metric into a distribution of distances from that ideal value.
Then, the complementary cumulative distribution function (CCDF) is used to transform the quantity by changing its distribution to be uniform and forcing it into the range $\interval{0}{1}$.
This can be understood as the first step of the probability integral transform, giving the original quantity a uniform distribution.
The CCDF expresses the cumulative probability to find a score that is less than the given metric, which here means the probability to find a metric value with a higher distance from the used ideal value.
In other words, the score is high for low distances and low for high distances from the ideal value.

This approach to transforming quantities to scores is applicable to arbitrary numeric quantities and not limited to software metrics.
The transformation using an ideal value can be understood as the generalization of a previous approach by \citeauthor{ulan2021copula}~\cite{ulan2021copula}.
In our work, we examined software metrics from the ``Qualitas.class corpus,'' which holds $23$ different types of software metrics for $111$ applications that are spread across eleven application domains~\cite{terra2013qualitas,honel2023qcc}.
The to-score transformation would be useless if raw quantities such as software metrics were distributed uniformly.
None of the metrics in the corpus follows a uniform distribution, meaning that the raw metrics cannot be compared, understood, or aggregated in a straightforward manner.
Applying the to-score transformation to a uniformly distributed quantity, however, would not have a negative impact.
Therefore, we recommend always applying the transform.
We also find that metrics are often\,---\,some metrics always\,---\,statistically significantly different across application domains.
For example, the metrics \emph{Depth Of Inheritance Tree}, \emph{Nested Block Depth}, or \emph{Cyclomatic Complexity} are commonly used in software quality models and we find them to always be different.
That means, that these metrics must be used with utmost care if the quality of the software product is to be determined by some mathematical aggregation.
Looking at the same problem from an application domain perspective unveils that the contexts in their individual entirety (i.e., including all metrics) are considerably different from each other.
That means that an application needs to be validated against a quality model that solely uses metrics from the application's target domain as otherwise, the results may be strongly misleading.

\begin{figure}[t!]
    \centering
    %\makebox[\textwidth][c]{\includegraphics[width=1.4\textwidth]{images/ac-proc.png}}
    \makebox[\textwidth][c]{\input{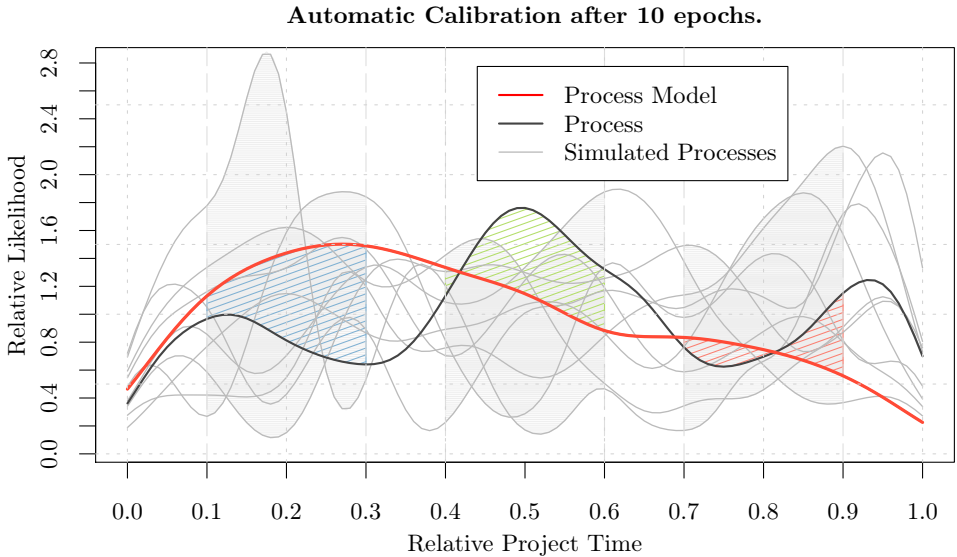}}
    \caption[Automatic Calibration of a process model using simulated random processes.]{{\small Random processes (light gray lines) are simulated to obtain marginal probability distributions for each defined deviation (blue, green, and red areas), which, in turn, are used for rank normalization/transformation.}}% The distributions are then used to transform the deviations between the process model (red line) and the newly observed process (black line) into scores.}}
    %\caption{{\small The process of Automatic Calibration, where the process model \textsf{REQ} and deviations are calibrated using random processes.}}
    \label{fig:ac-proc}
\end{figure}

\subsection{Interactive Transformations}\label{ssec:summary-honel2023masjoss}
\paragraph{Summary of~\citearticle{honel2023masjoss}.}
Many of the initial results of~\citepaper{honel2022qrsmas} were found by using the interactive application ``Metrics As Scores''~\cite{honel2023masgithub}.
%~\cite{honel2022masjoss,honel2023masgithub}.
While initially it was created to examine the Qualitas.class corpus, it has been adapted to work with own or arbitrary datasets, such as the well-known Iris flower- and Diamonds datasets~\cite{fisher1936iris,honel2023iris,honel2023diamonds}.
The interactive application can be thought of as a multiple analysis of variance~\cite{chambers2017statistical}.
While it also supports a large variety of statistical tests and the fitting of more than $\approx\utight{120}$ parametric distributions, its main goal is to enable the visual exploration of probability densities, cumulative distributions, scores (in terms of CCDFs), and quantile functions for all quantities across all the dataset's contexts.
Next to that, the application supports the automated creation of summary statistics and the generation of scientific reports that allow detailed insights into the differences and similarities among the studied contexts.

\section{Automatic Calibration}\label{sec:ac}
In order to transform an arbitrary numeric quantity into a score, we have established that this exercise requires a set of previously observed, context-specific values, as well as an ideal value.
Problem \textbf{P-VII} arose from the desire to minimize the qualitative effort required to evaluate a past software process in order to understand how severely affected it is by a problem such as the Fire Drill anti-pattern.
In \emph{Automatic Calibration}, we implement the solution of \textbf{P-VI} and apply the to-score transformation to deviations and divergences as previously observed between continuous\allowbreak
-time processes (continuous probability distributions of the temporal accumulation of each activity).
Simulating a large number of potential probability distributions allows us to approximate marginal distributions and optionally practical ideal values for each defined deviation or divergence.

\vspace{10pt}
\begin{mdframed}
    %\begin{center}
        For the remainder of this chapter, consider the following example which is also further illustrated by the figures and tables in this chapter.
    %\end{center}
\end{mdframed}

\paragraph{Example.}
Recall that we suggest modeling projects' activities using continuous\allowbreak%
-time processes.
Having observed a number of projects affected by the Fire Drill allowed us to compose weighted mixtures for each activity (see Section~\ref{ssec:obj2-quant-data}).
Each of these weighted mixtures represents a \emph{typical} accumulation for a given activity.
Suppose we were to observe many more affected projects and incorporate them into these existing mixtures according to their weight (severity).
Then, the mixtures would become sufficiently stable to serve as a \emph{continuous process model} (\textsf{PM}).
Having this process model and a new, previously unseen project (the process, \textsf{P}) at hand, we would use a previously trained regression model to predict the severity (see Section~\ref{sec:obj2-predict}).
Technically, this would be a curve comparison.
Using multiple measurement points, we would quantify where and how exactly two curves deviate from each other.
The regression model would use these \emph{raw features} as inputs.
Now suppose that the predicted severity is interesting enough to warrant an otherwise qualitative evaluation in order to find out how the concrete decision was arrived at (since the predictive model is a black box).
Then in order to reduce the qualitative efforts, Automatic Calibration suggests transforming each of the raw features into scores.
After calibration, the task of comparing curves becomes trivial, as scores can be compared in an apples-to-apples manner.
Furthermore, inspecting individual scores as per the result $\divergence{\mathsf{PM}}{\mathsf{P}}$ allows to get a simplified understanding of how strong the deviations actually are.

\subsection{Calibration}
\begin{figure}[!t]
    \makebox[\textwidth][c]{\input{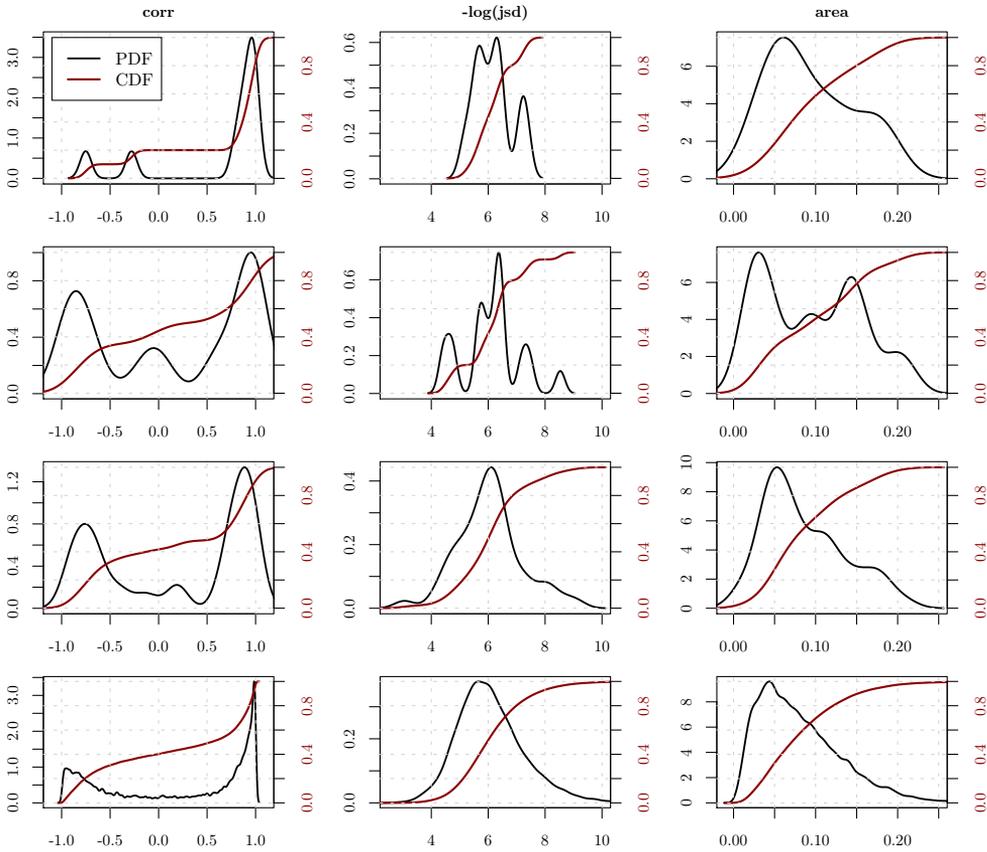}}
    \caption[Approximation of the kernel density estimates for the rank transform used in Automatic Calibration.]{{\small Approximation of the kernel density estimates for the rank transform used in Automatic Calibration after $n\tight{=}\set{10,20,50,10^4}$ epochs (top to bottom). Gaussian kernels and the Sheather--Jones rule for estimating the bandwidth are used. The left axis is the relative likelihood of the PDF and the right axis is the cumulative probability of the CDF.}}
    \label{fig:ac-approx}
\end{figure}

An exemplary calibration process is depicted in Figures~\ref{fig:ac-proc} and~\ref{fig:ac-approx}.
Here, we have taken the weighted mixture for the requirements-activity found in issue-tracking data as the process model \textsf{PM}.
Three different deviations are measured:
\begin{enumerate}[itemsep=.5ex]
    \item $\operatorname{\mathbf{corr}}$: Pearson sample correlation coefficient on the segment $\interval{0.1}{0.3}$ \cite{pearson1895corr} \cite[p.~116]{honel2023fdtr}. The coefficient has a range of $\interval{-1}{1}$, where $1$ means a strong positive linear correlation and is considered to be the best achievable value (the \emph{utopian} ideal value). $0$ means no correlation and we consider negative correlations to be even worse (that is, $-1$ is the worst possible value).
    \item $\operatorname{\mathbf{jsd}}$: Jensen--Shannon divergence on the segment $\interval{0.4}{0.6}$~\cite{Menendez1997jsd}. This divergence is symmetric (commutative) and measures the similarity between two probability distributions. Since we take the negative logarithm of it, its utopian ideal value is $\infty$. However, for reasons of numerical stability, we assume that the smallest observed divergence during calibration is the practically lowest-achievable divergence and use that as the ideal value, i.e., $i_{\operatorname{jsd}}\tight{=}\operatorname{\mathbf{sup}}\left[-\log{\left(X\right)}\right]$, where $X$ is the set of observed divergences.
    \item $\operatorname{\mathbf{area}}$: Absolute (total) area between curves, $\int\,\abs{\mathsf{PM}(x)-\mathsf{P}(x)}\,dx$, on the segment $\interval{0.7}{0.9}$. Since the theoretically lowest possible area between two lines or curves is $0$, we define $i_{\operatorname{area}}\tight{=}0$.
\end{enumerate}

The calibration itself requires \emph{usual} (context-specific/-dependent) values.
In order to gather these, we simulate a large number of processes as they \emph{might} occur (that is, we have not incorporated any further expectations into the simulation at this point).
For each simulated process, we calculate the three individual deviations and store the result.
After having observed sufficiently many random processes (and deviations), we can precisely approximate the empirical density and cumulative distribution function for each of the three features individually.
Instead of step-wise empirical cumulative distribution functions, we derive a smooth, closed-form CCDF from the kernel density estimate using Gaussian kernels (Equation~\ref{eq:kde}).
The bandwidth parameter $h$ (Equation~\ref{eq:kernel-bw}) is estimated using the Sheather--Jones (``SJ'') rule~\cite{sj1991bandwidth}.
For example, suppose the random variable $X$ holds all the observed areas.
The utopian ideal value $i_{\operatorname{area}}$ for the absolute area between two curves on any segment is zero (coincident lines).
The complementary random variable $D\tight{=}\abs{X-i_{\operatorname{area}}}$.
Since the areas are defined to be greater than or equal to zero and $i_{\operatorname{area}}\tight{=}0$, $X=D$.
The associated complementary CDF $\bar{F}_{\operatorname{area}}(x)$ is defined as in Equation~\eqref{eq:kde-ccdf}.
In cases where the ideal value $\utight{\neq}0$ and/or the deviation/divergence may assume arbitrary real values, we use the formulation of Equation~\eqref{eq:kde} to achieve the same result.

\begin{align}
%1 / length(temp) * sum(pnorm((x - temp) / h))
    \Phi(x)=&\;\frac{1}{2}\left(1+\operatorname{erf}\left(2^{-\sfrac{1}{2}}\,x\right)\right)\text{, the standard normal CDF,}\nonumber
    \\[1ex]
    N=&\;\operatorname{\mathbf{card}}\left(X\right)\text{, number of observations in $X$,}\nonumber
    \\[1ex]
    h\,\ldots&\;\text{Kernel bandwidth,}\label{eq:kernel-bw}
    \\[1ex]
    \bar{F}_{\operatorname{area}}(x)=&\;1-\frac{1}{N}\,\sum_{j=1}^{N}\,\Phi\left(\frac{\abs{x-i_{\operatorname{area}}}-D_j}{h}\right),\label{eq:kde}
    \\[0ex]
    &\;\text{using}\;X\tight{=}D\;\text{and}\;i_{\operatorname{area}}\tight{=}0\nonumber
    \\[1ex]
    =&\;1-\frac{1}{N}\,\sum_{i=1}^{N}\,\Phi\left(\left(x-X_{i}\right)\,h^{-1}\right).\label{eq:kde-ccdf}
\end{align}

Figure~\ref{fig:ac-approx} shows the approximated probability density (black) and cumulative distribution function (red) using Gaussian kernels for each of the three defined deviations.
The approximations are shown after $n\tight{=}\set{10,20,50,10^4}$ random processes.
With each observed deviation, the approximation gets more precise.
The differences between $50$ and $10^4$ simulated random processes are less significant than those between, e.g., ten and $20$ observations.
This suggests that Automatic Calibration requires comparatively few observations for a sufficiently accurate to-score transformation.

\subsection{Comparing Scores}
\begin{figure}[t!]
    \centering
    \makebox[\textwidth][c]{\input{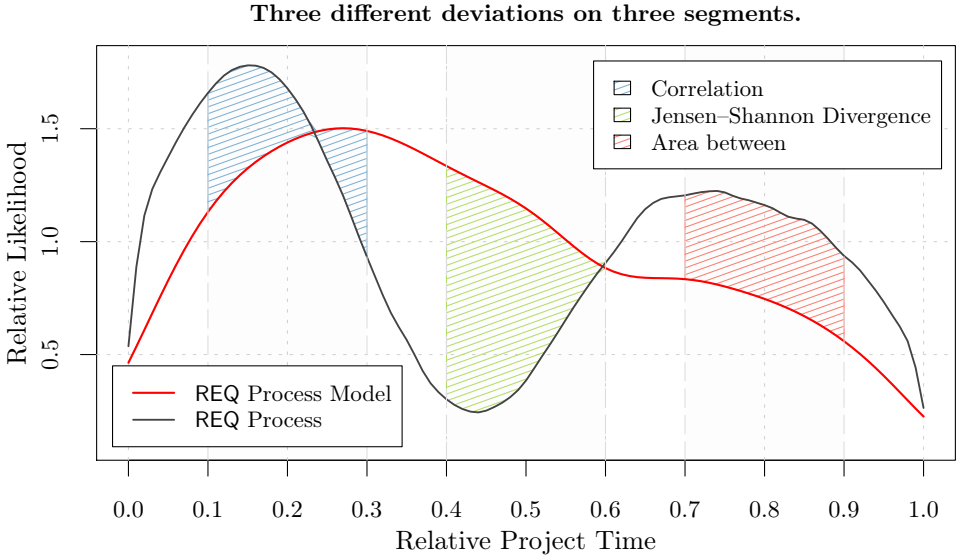}}
    \caption[Three different deviations between two continuous software processes.]{{\small The deviations between the weighted mixture (red) and a project (black) for the activity \textsf{REQ}.}}
    \label{fig:ac-raw-vs-scores}
\end{figure}

Having previously calibrated the raw features and derived scores, allows us now to compare them to the raw or z-standardized features.
Figure~\ref{fig:ac-raw-vs-scores} shows the three deviations that are computed in this chapter's example.
The \textsf{REQ} activity shown was taken from a project that has a ground truth of one out of ten (scaled to $0.1$).
Table~\ref{tab:ac-raw-vs-scores} shows the values of the raw features, the scaled features, the corresponding score, and the weighted score (using relative importance).
It becomes apparent now that the first two are of no value for the purpose of interpreting whether or not the two curves constitute a good match or not (they cannot \emph{explain} the ground truth).
We observe low and very low scores for this particular curve comparison.
Especially the Jensen--Shannon divergence does score particularly low, which is no surprise given how differently the two curves behave on segment $\interval{0.4}{0.6}$ and how important this score is.
The variable importance was determined by training a random forest.
While the average score is $\utight{\approx}0.239$, the dot product of the variable importance with the scores is $\utight{\approx}0.155$ and the random forest predicted a value of $\utight{\approx}0.08$, which all seem to be reasonable values given this toy example.
Note that the random forest was trained using leave-one-out, so its empirical estimate of the generalization error has a slightly pessimistic bias (i.e., it is likely to be lower in reality)~\cite{cawley2010overfitting}.

\begin{table}[t!]
    \centering
    %\resizebox{\textwidth}{!}{%
    \begin{tabular}{l|r|r|r}
            & $\operatorname{\mathbf{corr}}_{\interval{0.1}{0.3}}$ & $\operatorname{\mathbf{jsd}}_{\interval{0.4}{0.6}}$ & $\operatorname{\mathbf{area}}_{\interval{0.7}{0.9}}$ \\ \hline
        Raw Feature     & $-0.59459$    & $0.01749$  & $0.08280$ \\
        Z-Standardized  & $-1.49961$    & $3.75691$  & $2.57750$ \\
        As Score        & $0.28010$     & $0.02479$  & $0.41277$ \\
        Weighted Score  & $0.08428$     & $0.01395$  & $0.05638$ \\ \hline
        Relative Importance & $30.09$\% & $56.25$\% & $13.66$\% \\
    \end{tabular}
    %}
    \caption[Comparison of raw features against z-standardized and rank-transformed features.]{{\small Comparing raw features against z-standardized and rank-transformed features for the example shown in Figure~\ref{fig:ac-proc}.}}
    \label{tab:ac-raw-vs-scores}
\end{table}

A unit change is equal across any two scores that were previously calibrated using Automatic Calibration.
Therefore, we can also compare the achieved scores even though their underlying quantities are vastly different.
The correlation on the first segment achieved a score of $\utight{\approx}0.28$ out of $1.0$.
This seems reasonable when inspecting Figure~\ref{fig:ac-raw-vs-scores}, where only a smaller portion of the curves has a positive correlation.
The Jensen--Shannon divergence on the second segment scored very low with $\utight{\approx}0.02$.
Given that the two curves on that segment not only behave completely differently but also have a large spatial distance, a large divergence (and hence a low score) appears to be sound.
Lastly, the area in between curves scored $\utight{\approx}0.41$ on the last segment.
We observe a somewhat large area between the curves and this score tells us that more than $40$\% of the other simulated curves' area on that segment is even larger.

The achieved scores can be ordered as $\mathbf{area}>\mathbf{corr}>\mathbf{jsd}$ and their relative importance can be used to calculate an approximate result (total match).
However, only individual scores should be evaluated in order to replace portions of the qualitative evaluation, as a simple aggregation of scores might be misleading as it cannot capture more complex relations and non-linearities between scores and the dependent variable (even a simple aggregation is a kind of regression model, too, and it is likely to be different from the trained model).

When evaluating individual scores in order to find the smallest or largest deviations, we may also use the (scaled) variable importance to produce weighted scores, in order to better understand the outcome of the regression model (see Table~\ref{tab:ac-raw-vs-scores}).
The order for the scaled scores changes to $\mathbf{corr}>\mathbf{area}>\mathbf{jsd}$.
While we still have the lowest score/largest deviation/biggest problem with the divergence in the middle segment, the second-largest problem appears now to be the area between curves in the last segment.
It is these quantitatively derived insights that we suggest can add to the picture of qualitative evaluation.

\cleardoublepage
\chapter{Concluding Remarks and Future Work}\label{chap:conclusions-future-work}%
\vspace{5pt}
\chaptertoc
\chaptermark{Conclusions \& Future work}

\noindent
This chapter concludes the results of the presented publications and discusses their limitations.
This is followed by an outline of prospective future work.

\section{Conclusions}
In this dissertation, we suggest a path toward automated and accessible organizational learning.
It is largely based on capturing software evolution.
We propose three consecutive objectives with related problems.
For each problem, one or more solutions are suggested and the results of related publications are laid out.

The first objective,~\textbf{O-I}, seeks to accurately capture software size, evolution, and the nature of these changes.
As a solution to problem~\textbf{P-I}, we suggest the new software metric called source code density~\citepaperp{honel2018changeset}.
We find that it is a computationally cheap and largely language-independent metric.
Our studying of the source code density also reveals insights into developers' behavioral patterns, such as \emph{bulk-committing} and \emph{tangled changes}.
The second problem,~\textbf{P-II}, requires us to obtain a reliable, accurate, and generalizable classifier for changes.
Using the source code density and other size-based software metrics, we improve the previous state-of-the-art in commit classification~\cite{levin17boosting} considerably~\citepaperp{honel19importance}.
It allowed us to predict maintenance activities~\cite{swanson76acp} with great confidence.
In~\citearticle{honel2020using}, we propose an extension to our approach that also considers preceding commits.
This extension allows for another significant improvement in classification accuracy.
Lastly,~\citepaper{honel2023hmm} explores commit classification through a variety of more suitable models.
We demonstrate that commits do not occur at random and that joint conditional density models outperform other, state-of-the-art models.

The second objective,~\textbf{O-II}, seeks to harness application lifecycle management data for aspects of software process quality and organizational learning.
It poses three separate problems.
Problem~\textbf{P-III} requires us to identify if, which, and how application lifecycle management data can be used for process modeling.
Firstly, we approach this problem exploratively in~\citepaper{picha_honel22pilot}.
Similar to the Unified Process~\cite{Scott2001uniproc}, we suggest modeling activities from source code and issue-tracking data.
Visualization of such activities alone reveals discernible patterns of problems in affected projects.
However, based on our findings, we revise our approach to modeling activities and are able to suggest a uniform way to do so in~\citearticle{honel2023embedded}.
Modeling processes of problems using continuous-time accumulations of maintenance activities and issue-tracking reports yields human-comprehensible representations.
However, for problem~\textbf{P-IV}, we desire to find a predictive model for assessing if and to what degree a problem is present in such a process.
While in~\citepaper{picha_honel22pilot} we would first attempt to employ expert-engineered binary decision rules, our validation shows that these perform poorly, as they cannot capture the many facets of problems that manifest as complex phenomena.
Therefore, we suggest an adaptive training workflow in~\citearticle{honel2023embedded}.
It is designed to obtain precise and robust estimates and confidence intervals for the generalization error of predictive models that are trained to assess the severity of the manifestation of a problem, even with ``little data''.
It allowed us to obtain models that can predict the severity with acceptable generalization error trained on a few ($\utight{\approx}\mathbf{12\tight{-}20}$) projects.
The last problem of the second objective,~\textbf{P-V}, aims at finding a replicable methodological framework for automatic organizational learning from only a few past projects.
Effective training on scarce data by means of the adaptive training workflow is part of the solution.
However, this workflow requires a reliable ground truth.
In~\citearticle{honel2023embedded}, therefore, we further suggest how a ground truth mined by independent experts can be quality-assured by means of inter-rater agreement.
We further allow the framework to generalize to potentially many other activity-based phenomena and contexts and encourage its replication with these.

In the last objective,~\textbf{O-III}, we seek to enhance the explainability of quantitative data and to reduce the amount of required qualitative analysis for evaluating past projects.
Since concepts such as explainability and understandability are often context-dependent~\cite{zagzebski2017whatis}, we limit our focus by wanting to enhance these by demystifying the input data and aggregations thereof in black-box models.
Derived from that, we define problem~\textbf{P-VI} as a lack of a technique for transforming raw quantities into something that can be understood, compared, and associated with quality.
In~\citepaper{honel2022qrsmas}, we argue that software metrics in particular suffer from a lack of interpretability, especially when inspected out of context~\cite{gil2016meannothing}.
We suggest a contextual approach that can automatically derive typical values for software metrics and use these to transform them into distances.
This concept called \emph{score} provides a solution to each individual facet of the posed problem:
Any two scores can be compared, a score can be associated with quality (higher is better), all scores have the same range $\interval{0}{1}$, and the same unit change across any two scores can be considered equally good/bad.
\citearticle{honel2023masjoss} introduces a tool- and (visual) analysis suite that implements this approach.
The last problem,~\textbf{P-VII}, suggests that we shall now combine the solutions to the previous problems in order to find a data-based solution for evaluating complex phenomena that can replace aspects of otherwise qualitative evaluation.
We contemplate an approach called \emph{Automatic Calibration} and demonstrate its feasibility using some preliminary results.
This approach relies on all previously introduced solutions, from capturing software evolution using software size, source code density, and other suitable application lifecycle management data, modeling of maintenance activities, to transforming software metrics to scores.
It uses large numbers of random processes to calibrate scores and combines the result with variable importance.
These together can then be facilitated for explaining decisions of black-box models, such as Random forests, and\,---\,at least partially\,---\,replace qualitative evaluation.

\section{Limitations}
While application lifecycle management offers a large variety of exploitable artifacts, we limit ourselves to data from version control systems and issue-tracking applications.
This choice came about naturally, as we are deriving maintenance activities from software size and how it changes during software evolution.
Early on we learned a limitation of software size, in that it cannot estimate effort or productivity well in conjunction with\,---\,admittedly\,---\,coarse notions of time spent.
However, there might be other sources of data that are perhaps more or less suitable for organizational learning than the data we chose.
The activities we chose to model do only partially overlap with the \emph{disciplines} as defined in some flavors of the Unified Process~\cite{Scott2001uniproc,kruchten2003rup}.
Still, we accomplish the task of detecting the presence and assessing the severity of complex phenomena like the Fire Drill, post mortem.
We have argued for the applicability of this approach to other, similar, activity-based phenomena.
However, the suitability can only be determined by carrying out subsequent case studies with modified contexts, that is, for example, by modeling different activities, studying other phenomena, or by altering the boundary conditions (e.g., academic vs. industrial or examining non-agile settings).
Our approach to enhancing the understandability of quantitative data and -models is a combination of using scores and variable importance.
While it can certainly help to explain some of the results of the otherwise latent predictive model, it can only partially replace qualitative evaluation.
Furthermore, there likely exist a large number of potential alternative approaches that would be suitable for enhancing the comprehensibility of application lifecycle management data and models thereof~\cite{burkart2021explain}.

\section{Future Work}
We want to outline two encouraging branches of future work in particular.
The first branch is that of subsequent replications of the framework suggested in our embedded case study~\citearticlep{honel2023embedded}.
Our empirical observations are reasonably valid for the studied case and context and we argue for the validity of the suggested framework in similar scenarios.
However, only subsequent (partially) replicating case studies can confirm or reject our hypothesis.
Replication of the embedded case study will likely yield some other interesting insights, depending on how the case and context are altered.
For example, if we were to replicate the exact same study in the future, we could speak to the validity of the original results.
If, for example, we would study one of the other suggested phenomena (such as ``Collective Procrastination''~\cite{picha2019apdetect} or ``Brook's Law''~\cite{brooks1995myth}) in the same context, we would be able to learn about the aptness of the activities as chosen earlier, or whether source code data works better than issue-tracking data, etc.
Within the confines of the studied context, the projects were forced to implement solutions that involved descoping, the only valid solution for a Fire Drill in this case.
Therefore, it would be interesting how affected projects deal with this difficulty in, e.g., an industrial setting.

The second branch is that of using scores in multi-objective optimization scenarios.
%We have previously demonstrated the usefulness of scores for accessible organizational learning from quantitative data.
In organizational settings, the task of multiple-criteria decision-making (MCDM) often arises.
Making decisions based on multiple or many criteria is non-trivial and requires careful consideration of all criteria and a well-structured problem~\cite{franco2011mcdm}.
As a specialized area of MCDM, multi-objective optimization is concerned with the mathematical optimization of two or more conflicting objectives and how to minimize them simultaneously~\cite{miettinen1998}.
This is shown in Equation~\eqref{eq:moop}.
The goal is to find some \emph{Pareto efficient} solution $\mathbf{x}^\star$ in the \emph{decision space} $\mathcal{D}$ that minimizes the loss of all objectives simultaneously such that no single objective can be further improved without worsening another one.
\begin{align}
    \min_{\mathbf{x}\in\mathcal{D}}\,\left\{f_1(\mathbf{x}),\dots,f_k(\mathbf{x})\right\},\text{ where}\;k\geq2~\label{eq:moop}
\end{align}
The continuous nature of multi-objective optimization implies that an infinite number of solutions exist and that the feasible alternatives are not known in advance.
The Pareto efficient set contains those solutions that cannot be further improved.
However, each solution represents a unique trade-off and all solutions are, in a mathematical sense, considered to be equally good~\cite{branke2008moo}.
The lack of orderability requires consulting a human decision-maker and applying one of the four philosophies of no-preference methods, a priori methods, a posteriori methods, and interactive methods~\cite{hwang1979modm}.
In an a priori method, a decision-maker specifies preference information before the solution process starts~\cite{miettinen2008}.
However, the availability and feasibility of the desired solution are most likely not available beforehand.
Worse, the use of weights for expressing preference often is misleading~\cite{bernard1996}.

By using scores instead of raw objectives, we have previously demonstrated an improvement in finding solutions in multi-objective optimization when preference is expressed a priori or the correctness of the found trade-offs is to be assessed a posteriori~\cite{honel2022pareto}. 
%(see Appendix~\ref{app:publ-honel2022pareto}).
We show that the solution space must be thought of as non-homogeneous.
A multi-objective problem arises, for example, when we want to align two processes (e.g., updating a process model using one or more observed processes) using two or more conflicting objectives.
We have previously demonstrated that this is a non-trivial task that cannot be accomplished without the usage of scores because a decision-maker's preference will almost surely not lead to the desired trade-off (the interested reader may regard an illustrative and comprehensive example that attempts to align continuous processes using scores as shown in~\cite[pp.~41--168]{honel2023fdtr}).

\cleardoublepage

%----------------------------------------------------------------------------------------------------
%---------------------HERE STARTS THE APPENDIX -----------------------------------
%----------------------------------------------------------------------------------------------------

\phantomsection
\addcontentsline{toc}{chapter}{Bibliography}

\renewcommand*{\bibfont}{\small}
\printbibliography
\cleardoublepage

\begin{appendices}
%\appendixpage
\noappendicestocpagenum
\addappheadtotoc

\renewcommand{\myfullcite}[1]{\printpublication{#1}.}

\renewcommand{\appendixname}{Paper}
\chapter{A changeset-based approach to assess source code density and developer efficacy}\label{app:publ-honel2018changeset}
\chaptermark{\label{app:publ-honel2018changeset}}
\myfullcite{honel2018changeset}

\chapter{Importance and Aptitude of Source Code Density for Commit Classification into Maintenance Activities}\label{app:publ-honel19importance}
\chaptermark{\label{app:publ-honel19importance}}
\myfullcite{honel19importance}

\renewcommand{\appendixname}{Article}
\chapter{Using source code density to improve the accuracy of automatic commit classification into maintenance activities}\label{app:publ-honel2020using}
\chaptermark{\label{app:publ-honel2020using}}
\myfullcite{honel2020using}

\renewcommand{\appendixname}{Paper}
\chapter{Exploiting Relations, Sojourn-Times, and Joint Conditional Probabilities for Automated Commit Classification}\label{app:publ-honel2023hmm}
\chaptermark{\label{app:publ-honel2023hmm}}
\myfullcite{honel2023hmm}

\chapter{\texorpdfstring{Process Anti-Pattern Detection\,--\,a Case Study}{Process Anti-Pattern Detection - a Case Study}}\label{app:publ-picha_honel22pilot}
\chaptermark{\label{app:publ-picha_honel22pilot}}
\myfullcite{picha_honel22pilot}

\renewcommand{\appendixname}{Article}
\chapter{Activity-Based Detection of (Anti-)Patterns: An Embedded Case Study of the Fire Drill}\label{app:publ-honel2023embedded}
\chaptermark{\label{app:publ-honel2023embedded}}
\myfullcite{honel2023embedded}

\renewcommand{\appendixname}{Paper}
\chapter{Contextual Operationalization of Metrics as Scores: Is My Metric Value Good?}\label{app:publ-honel2022qrsmas}
\chaptermark{\label{app:publ-honel2022qrsmas}}
\myfullcite{honel2022qrsmas}

\renewcommand{\appendixname}{Article}
\chapter{Metrics As Scores: A Tool- and Analysis Suite and Interactive Application for Exploring Context-Dependent Distributions}\label{app:publ-honel2023masjoss}
\chaptermark{\label{app:publ-honel2023masjoss}}
\myfullcite{honel2023masjoss}

\end{appendices}
\cleardoublepage



\section*{Supplementary material}

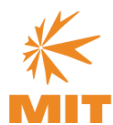

# The Swedish Research School of Management and Information Technology MIT

The *Swedish Research School of Management and Information Technology* (MIT) is one of the 16 national research schools supported by the Swedish Government. MIT is jointly operated by the following institutions: Blekinge Institute of Technology, Chalmers University of Technology, University of Gothenburg, Jönköping International Business School, Karlstad University, Linköping University, Linnaeus University Växjö, Lund University, Mälardalen University, Stockholm University, Umeå University, Örebro University, and Uppsala University, host to the research school. At the Swedish Research School of Management and Information Technology (MIT), research is conducted, and doctoral education provided, in three fields: management information systems, business administration, and informatics.


# DISSERTATIONS FROM THE SWEDISH RESEARCH SCHOOL OF MANAGEMENT AND INFORMATION TECHNOLOGY

## Doctoral theses (2003- )

1. **Baraldi, Enrico (2003),** *When Information Technology Faces Resource Interaction: Using IT Tools to Handle Products at IKEA and Edsbyn.* Department of Business Studies, Uppsala University, Doctoral Thesis.

2. **Wang, Zhiping (2004),** *Capacity-Constrained Production-Inventory Systems: Modelling and Analysis in both a Traditional and an E-Business Context.* IDA-EIS, Linköpings universitet och Tekniska Högskolan i Linköping, Doctoral Thesis.

3. **Ekman, Peter (2006),** *Enterprise Systems & Business Relationships: The Utilization of IT in the Business with Customers and Suppliers.* School of Business, Mälardalen University, Doctoral Dissertation.

4. **Lindh, Cecilia (2006),** *Business Relationships and Integration of Information Technology.* School of Business, Mälardalen University, Doctoral Dissertation.

5. **Frimanson, Lars (2006),** *Management Accounting and Business Relationships from a Supplier Perspective.* Department of Business Studies, Uppsala University, Doctoral Thesis.

6. **Johansson, Niklas (2007),** *Self-Service Recovery.* Information Systems, Faculty of Economic Sciences, Communication and IT, Karlstad University, Doctoral Thesis.

7. **Sonesson, Olle (2007),** *Tjänsteutveckling med personal medverkan: En studie av banktjänster.* Företagsekonomi, Fakulteten för ekonomi, kommunikation och IT, Karlstads universitet, Doktorsavhandling.

8. **Maaninen-Olsson, Eva (2007),** *Projekt i tid och rum: Kunskapsintegrering mellan projektet och dess historiska och organisatoriska kontext.* Företagsekonomiska institutionen, Uppsala universitet, Doctoral Thesis.


\cleardoublepage

\backmatter
\cleardoublepage

\end{document}